\documentclass[fleqn,usenatbib]{mnras}

\usepackage{newtxtext,newtxmath}
\usepackage[english,english]{babel}
\usepackage{amsmath}
\usepackage{amssymb,amsfonts,textcomp}
\usepackage{array}
\usepackage{supertabular}
\usepackage{hhline}
\usepackage[usenames]{color}
\usepackage{xspace}
\usepackage{tabularx}
\usepackage[T1]{fontenc}
\usepackage{ae,aecompl}
\usepackage[normalem]{ulem}

\usepackage{graphicx}	
\usepackage{amsmath}	
\usepackage{amssymb}	
\usepackage{caption}
\usepackage{rotating}
\usepackage{sidecap}
\usepackage{csvsimple}
\usepackage{color}
\usepackage{xcolor}
\usepackage{soul}

\def \apj {{\it ApJ}}
\def \apjl {{\it ApJ Let.}}
\def \apjs {{\it ApJ Sup.}}

\def \araa {{\it ARA\&A}}

\def \aap {{\it A\&A}}
\def \aj {{\it AJ}}
\def \mnras {{\it MNRAS}}
\def \nat {{\it Nature}}

\newcommand{\msun}{\rm M_{\odot}}

\newcommand{\HI}{H{\,\sc i}\xspace}
\newcommand{\HII}{H{\,\sc ii}\xspace}

\newcommand{\dfq}{\Delta f_q}
\newcommand{\dfD}{\Delta f_D}
\newcommand{\dfM}{\Delta f_M}
\newcommand{\fatm}{f_{\rm atm}}
\newcommand{\be}{\begin{equation}}
\newcommand{\ee}{\end{equation}}

\title[]{Angular momentum-related probe of cold gas deficiencies}

\author[J. Li et al.]{
Jie Li,$^{1}$\thanks{E-mail: jie.li@icrar.org}
Danail Obreschkow,$^{1,2}$
Claudia Lagos$^{1,2}$, Luca Cortese$^{1,2}$, Charlotte Welker$^{1}$, \newauthor{and Robert D\v{z}ud\v{z}ar$^{3}$}
\\
$^{1}$International Centre for Radio Astronomy Research, University of Western Australia, 7 Fairway, Perth, WA 6009, Australia\\
$^{2}$ARC Centre of Excellence for All Sky Astrophysics in 3 Dimensions (ASTRO 3D)\\
$^{3}$Centre for Astrophysics and Supercomputing, Swinburne University of Technology, PO Box 218, Hawthorn, VIC 3122, Australia
}

\date{Accepted XXX. Received YYY; in original form ZZZ}

\pubyear{2018}

\begin{document}
\label{firstpage}
\pagerange{\pageref{firstpage}--\pageref{lastpage}}
\maketitle

\begin{abstract}
Recent studies of neutral atomic hydrogen (\HI) in nearby galaxies found that all field disk galaxies are \HI saturated, in that they carry roughly as much \HI as permitted before this gas becomes gravitationally unstable. By taking this \HI saturation for granted, the atomic gas fraction $\fatm$ of galactic disks can be predicted as a function of the stability parameter $q=j\sigma/(GM)$, where $M$ and $j$ are the baryonic mass and specific angular momentum of the disk and $\sigma$ is the \HI velocity dispersion \citep{Obreschkow2016}. The log-ratio $\dfq$ between this predictor and the observed atomic fraction can be seen as a physically motivated `\HI deficiency'. While field disk galaxies have $\dfq\approx0$, objects subject to environmental removal of \HI are expected to have $\Delta f_q>0$. Within this framework, we revisit the \HI deficiencies of satellite galaxies in the Virgo cluster and in clusters of the EAGLE simulation. We find that observed and simulated cluster galaxies are \HI deficient and that $\dfq$ slightly increases when getting closer to the cluster centres. The $\dfq$ values are similar to traditional \HI deficiency estimators, but $\dfq$ is more directly comparable between observations and simulations than morphology-based deficiency estimators. By tracking the simulated \HI deficient cluster galaxies back in time, we confirm that $\Delta f_q\approx0$ until the galaxies first enter a halo with $M_{\rm halo}>10^{13}\msun$, at which moment they quickly lose \HI by environmental effects. Finally, we use the simulation to investigate the links between $\dfq$ and quenching of star formation.
\end{abstract}

\begin{keywords}
galaxies formation ---
galaxies evolution ---
galaxies interactions ---
galaxies kinematics and dynamics
\end{keywords}



\section{Introduction}
\label{intro}

Neutral atomic gas (\HI) is found in most galaxies. It is a pivotal way-station in the evolution of galaxies: cooling accretion flows deliver \HI gas to galactic disks \citep[e.g.][]{Voort2012,Faucher2015}, where, given the right conditions, the \HI can collapse into clouds and combine into molecular gas (H$_2$) and further into stars. In turn, \HI can also be the exhaust product of stellar winds and supernovae and get pushed out of galactic disks by energetic feedback \citep[e.g.][]{Faber1976,Faucher2015,Ford2013,Lagos2014}.

Because of the physical importance of \HI, the amount of this gas in a galaxy is a key parameter in galaxy evolution studies. However, since field disk galaxies span many ($>5$) orders of magnitude in stellar (and halo) mass, the most physically meaningful quantity is the atomic mass \textit{fraction}, rather than the absolute amount of \HI. In this work, the neutral atomic gas fraction of a galaxy is defined as
\be
	f_{\rm atm}\equiv \frac{1.35M_{\rm {HI}}}{M},
	\label{equ:fatm}
\ee
\noindent where $M=M_{*}+1.35(M_{\rm{HI}}+M_{\rm{H_2}})$ is the baryonic mass, $M_{*}$ is the stellar mass, $M_{\rm{HI}}$ is the \HI mass, $M_{\rm{H_2}}$ is the molecular hydrogen mass and the factor 1.35 accounts for the universal Helium fraction, which is hard to measure directly on a galaxy-by-galaxy basis.

Observations of \HI in rest-frame 21cm emission revealed that $f_{\rm atm}$ varies considerably between different galaxies \citep{Maddox2015,Catinella2018,Chung2009}. These empirical variations hold interesting clues on qualitatively different physics. However, a priori, it is not obvious whether these differences are due to \textit{internal} differences (i.e.\ differences in other galaxy properties) or to \textit{external} differences related to the galactic environment. In fact, both can be important. As for internal differences, it is well established that $\fatm$ exhibits a weak but systematic dependence on stellar mass \citep{Catinella2010} and a pronounced dependence on the morphology \citep{Haynes1984a}, size \citep{Boselli2009} and spin parameter (\citealp{Huang2012}) of the galactic disk. Even without an understanding of the causalities between these observables, it is thus clear that $\fatm$ relates strongly to internal physics. In turn, it is also well established that $\fatm$ can be strongly reduced by environmental effects, especially in dense cluster environments. Already early observations \citep[e.g.][]{Davies1973,Giovanelli1985} have found that spiral galaxies near the core region of clusters are very deficient in \HI, compared to field galaxies of similar morphology and size. Many mechanisms can drive such deficiencies \citep[see review][]{Boselli2014c}; for example (1) stripping of \HI by the ram pressure of the hot intra-cluster medium \citealp{Gunn1972}, (2) tidal stripping by gravitational forces \citep{Merritt1983}, (3) heating by dynamical friction causing the \HI to be `harassed' \citep{Moore1996}, (4) merger-driven starbursts consuming a lot of \HI \citep{Hopkins2006} and (5) suppressed supply of new gas despite ongoing star formation (strangulation) \citep{Peng2015}.

In studying environmental effects on \HI, it is crucial to separate the most important internal effects on $\fatm$ from external ones.  In other words, we would like to measure the deficiency (or excess) in $\fatm$ due to external effects. The definition of such a `deficiency' parameter is not obvious, as it requires to calibrate the typical \HI content of galaxies against a property that is not (or significantly less) affected by the environment  than the cold gas reservoir. From an observational point of view, the concept of \HI deficiency has originally been introduced by \citet{Haynes1984a}  as the difference, in terms of $\log(\fatm)$, between the observed \HI mass and the value expected for a field galaxy with the same morphological type and optical diameter. While empirical, this relation is based on the assumption that environment affects \HI without affecting the optical size. Admittedly, this definition has hampered a quantitative comparison with predictions from theoretical models for which visual morphologies cannot be easily obtained. Thus, in recent years, significant effort has gone into calibrating new \HI-deficiency parameters using physical quantities directly comparable with predictions from models. Among them, the combination of stellar surface density and ultraviolet colour presented by \citet{Catinella2010,Catinella2018}  is the one that mimics most closely the original definition, and has allowed detailed comparisons with semi-analytical models of ram-pressure and starvation \citep[see][]{Cortese2011}. Conversely, calibrations based on just stellar mass or optical luminosity perform quite poorly as mass is a poor predictor of $\fatm$ \citep[see][]{Catinella2010,Brown2015}.

An interesting alternative to an empirical calibration of $\fatm$ is offered by a recently proposed parameter-free physical model for $\fatm$ in disk galaxies \citep[][hereafter O16]{Obreschkow2016}. An empirical extension of this model to other baryonic components has recently been presented by \citet{Romeo2020} \citep[see also][]{Romeo2018}. The O16 model (detailed in Section~\ref{HIdeficiency}) relies on the assumption that disk galaxies (as defined in the next paragraph) contain as much \HI as they can in a gravitationally stable manner, given their baryonic mass $M$ and baryonic specific angular momentum (sAM) $j=J/M$. Apart from the straightforward physical interpretation of this calibration, the model of O16 has the advantage that it can be applied to observational data in the same way as to numerical simulations \citep[e.g.][]{Stevens2018a,Wang2018}, hence allowing for an unbiased comparison.

In this paper, we use the O16 model to revisit \HI deficiencies in the satellite galaxies of the Virgo cluster, studied in the VIVA survey \citep{Chung2009}, and compare these measurements to cluster galaxies in the EAGLE simulation \citep{Schaye2014,Crain2015,McAlpine2016}. The objective is to critically discuss the use of the O16 model as a diagnostics of environmental effects which affects primarily the gas fraction but not (or much less) $j$, as well as to use this model as a bridge between observations and simulations to better understand the nature and implications of such effects. 

Throughout this work, we will concentrate on `disk' galaxies, here defined as all objects showing a clear signature of global rotation in their \HI velocity maps. These galaxies encompass spiral and irregular optical morphologies and they can exhibit environmental disturbances such as tidal or pressure-driven stripping. For comparison with cluster galaxies, we will use the term `field' galaxies to denote objects, which are not part of a group/cluster, show no major companion and no signs of ongoing or past interactions/mergers in their \HI maps.

This paper is organised as follows. We start by summarising the O16 model and defining a \HI deficiency estimator based on this model (Section~\ref{HIdeficiency}). Section \ref{VIVA} explores this \HI deficiency in Virgo cluster galaxies and compares this deficiency to traditional estimators. Section \ref{EAGLE} explores the OG16-based \HI deficiency estimator in the EAGLE simulation and compares the results to observations. Conclusions and a summary are given in Section \ref{conclusions}.

\section{Physically motivated \HI deficiency}\label{HIdeficiency}

In an effort to explain the observed correlations between the \HI content and $j$ in field disk galaxies, O16 introduced an analytical model for the atomic gas mass that can be supported against gravitational collapse in a flat exponential disk with circular rotation. Relying on \citet{Toomre1964} like stability considerations, they found that the maximum stable value of $f_{\rm atm}$ depends on the mass and kinematics of a disk through a single dimensionless parameter $q$, defined as
\be\label{equ:q}
q=\frac{j\sigma}{GM},
\ee
where $G$ is the gravitational constant, $M$ and $j$ are the baryonic (stars+cold gas) mass and sAM and $\sigma$ is the one-dimensional velocity dispersion of the `warm' \HI gas. We take this dispersion to be $\sigma=10$~km/s, consistent with the observational results \citep{Walter2008,Leroy2008} that nearby spiral galaxies exhibit a galaxy-independent and radius-independent dispersion between roughly 8~km/s and 12~km/s. Incidentally, this dispersion approximately corresponds to the thermal motion of hydrogen and the supersonic turbulence in the interstellar medium at a temperature of $10^4\rm~K$, characteristic for the phase transition from \HII to \HI. To the extent that $\sigma$ is considered fixed by such fundamental physical considerations, the O16 model is completely parameter-free.

\begin{figure}
\includegraphics[width=1.0\columnwidth]{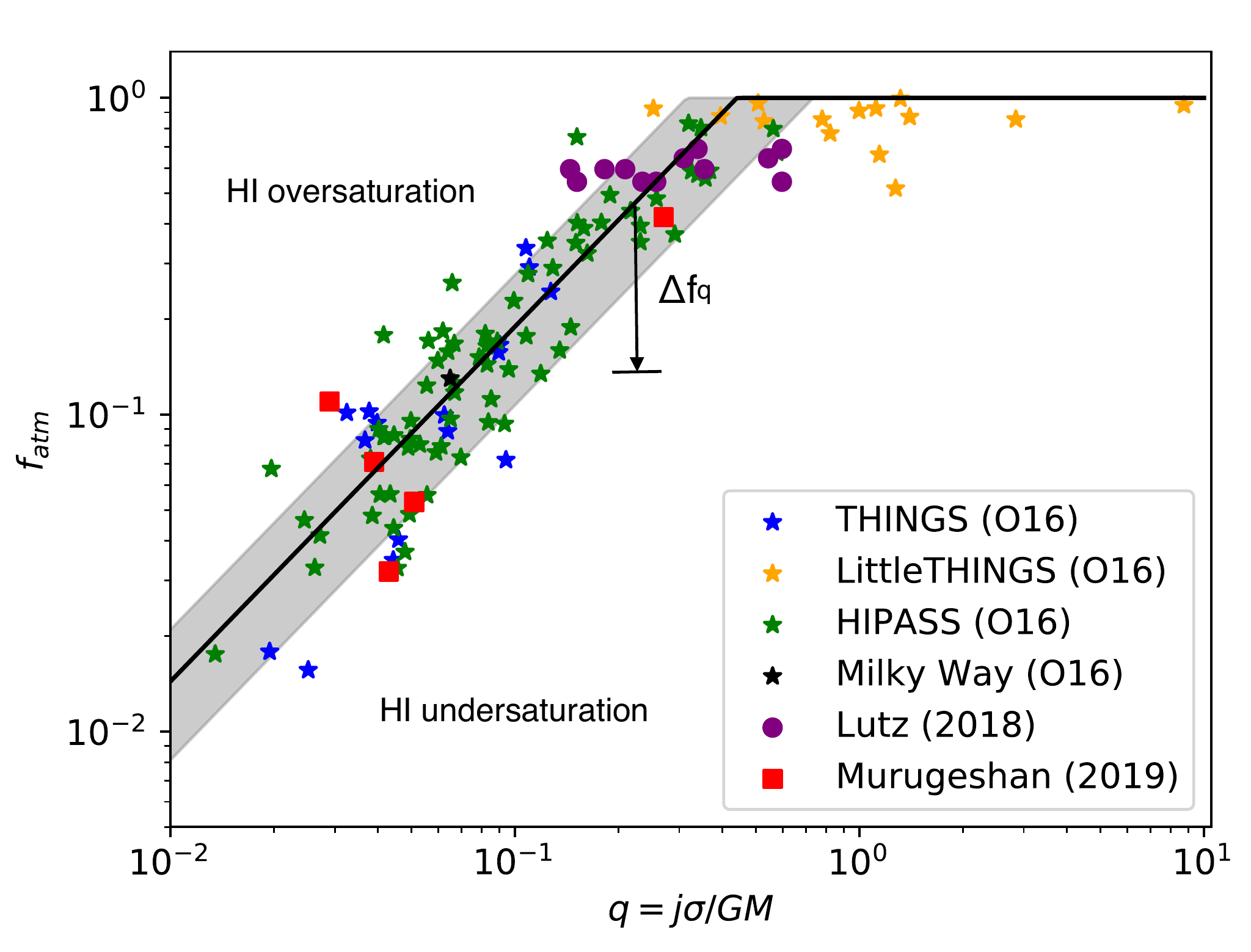}
\caption{Atomic gas fraction versus the $q$-parameter for local field galaxies from different samples. The black line is the approximation (Eq.~(\ref{equ:fq})) to the O16 model, with its uncertainty region due to the variance of the velocity dispersion shown as gray shading. }
\label{fig:definitionfq}
\end{figure}

The exact relation between $q$ and $f_{\rm atm}$ depends on the rotation velocity as a function of radius. However, this effect of the rotation curve is small (see different lines in Figure~2 of O16) and the $q$--$f_{\rm atm}$ relation is generally well approximated by the truncated power law

\begin{equation}
\label{equ:fq}
f_{\rm atm}={\rm min}\{1,2.5q^{1.12}\}.
\end{equation}
This function is plotted as the solid line in Figure~\ref{fig:definitionfq}.

All field disk galaxies (as defined in Section~1) with sufficient \HI and optical data for an accurate determination of $q$ and $f_{\rm atm}$, analysed so far, satisfy this relation within a log-normal scatter of about 0.16~dex standard deviation, as shown in Figure~\ref{fig:definitionfq}. The data shown here spans five orders of magnitude in stellar mass and include galaxies that have very high \HI fractions \citep{Lutz2018} and low \HI fractions \citep{Murugeshan2019} for their stellar mass and absolute $r$-band magnitude, respectively. In other words, these \HI extreme galaxies have extremely high/low sAM $j$ for their mass (see Figure~\ref{fig:Mfatm_j}), but for their effective stability parameter $q$, their \HI content is, in fact, normal. This result suggests that field disk galaxies in the local universe reside at the \HI saturation point, approximated by Eq.~(\ref{equ:fq}). Interestingly, a recent analysis of cosmological zoom-simulations \citep{Wang2018} showed that this statement holds true for most field  disk galaxies at any redshift. This implies that Eq.~(\ref{equ:fq}) is a nearly universal and physically motivated relation for field disk galaxies.

Given the accuracy of the O16 model in predicting the \HI mass of undisturbed galaxies, it seems sensible to define the \HI deficiency of galaxies by their offset from Eq.~(\ref{equ:fq}), i.e.~as
\be\label{equ:delfq}
\dfq={\rm log_{10}}\left({\rm min}\{1,2.5q^{1.12}\}\right)-{\rm log_{10}}\left(f_{\rm atm}\right).
\ee
The higher the value of $\dfq$, the more \HI deficient a galaxy is. 

With respect to more common definitions of the \HI deficiency (e.g.~see Section \ref{ss:otherestimators}), $\dfq$ has the advantages that it can be directly interpreted as the \HI deficiency relative to the saturation point and that it does not require any tuning to a reference sample. As we will demonstrate in the following, $\dfq$ can be accurately measured in kinematic observations and simulations and thus offers a new bridge for comparing them in a like-to-like fashion.

\section{Clues from the VIVA observation}
\label{VIVA}

\subsection{Sample and data}

To study the position of real cluster galaxies in the $(q,\fatm)$-plane, we rely on the data from the VLA Imaging of Virgo in Atomic gas (VIVA, \citealp{Chung2009}) survey. These data are optimal for this purpose because they provide us with spatially resolved 21cm spectral line data, showing both the amount of \HI and its Doppler velocity along the line-of-sight at a spatial resolution of 15" (1.2~kpc at 17~Mpc). The use of such resolved kinematic data permits us to avoid the usual assumptions that (1) most material orbits the galaxy at the maximum circular velocity and (2) that this velocity is measured by the 21cm line width (e.g.~$W_{50}$). In fact, both these assumptions, especially the second one, are hard to justify in stripped systems. The full VIVA sample counts 53 galaxies (48 spirals and 5 irregulars), all showing at least some rotation. In this sense and in terms of the range of stellar $j/M$ the VIVA sample is similar to the the THINGS sample.

Our analysis also requires stellar masses and optical sizes. These were drawn from the analysis of the GALEX-enhanced Herschel Reference Survey (HRS) data by \citet{Cortese2012}, where stellar masses were derived using a ($g$-$i$)-colour-dependent mass-to-light ratio. Of the 53 VIVA objects, 41 are given stellar masses. We deliberately exclude merging or strongly interacting systems, where two galaxies are visibly connected in \HI. These are four galaxies  (NGC 4294/4299, NGC 4567/4568), of which one was already rejected because it had no separate stellar mass. This leaves us with a sample of 38 galaxies for the present analysis.

The definitions of $\fatm$ and $q$ both refer to all baryons (stars+cold gas). Sometimes, the cold gas can have a significant molecular component. In the analysis of this paper, we include molecular masses determined from $^{12}$CO(1--0) emission in the cold gas study of the HRS \citep{Boselli2014b}. We used the data corresponding to a constant CO-to-H$_2$ conversion $X_{\rm CO}=2.3\cdot10^{20}\rm~cm^{-2}/(K~km\,s^{-1})$ -- the typical Milky Way value, not accounting for helium \citep{Bolatto2013}. Of our 38 galaxies, 37 have CO data. The molecular mass of the remaining object NGC 4606 is neglected.

All data are analysed assuming that the galaxies are situated at a distance of 17~Mpc, except for NGC~4380 and NGC~4424 which are assumed to lie at the Virgo~B distance of 23~Mpc and NGC~4561 at a Hubble flow distance of 20.14~Mpc (following the HRS data).

\begin{figure*}
\includegraphics[width=0.0828\textwidth]{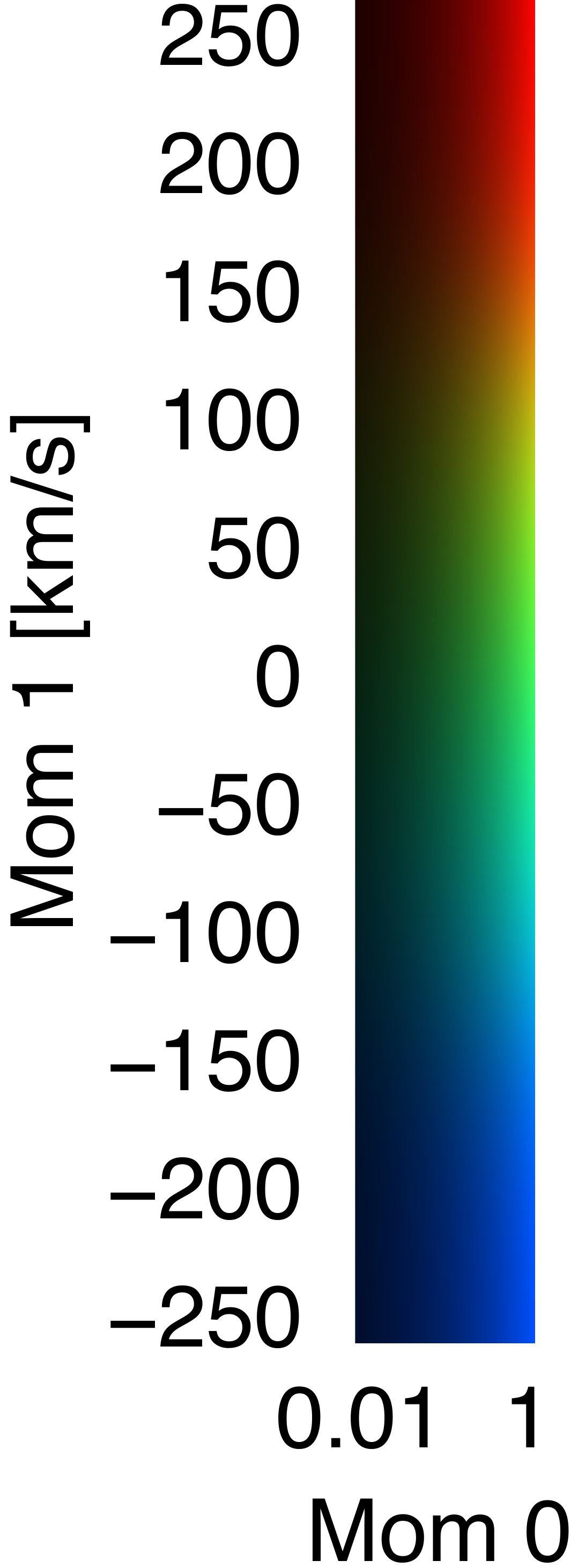}
\includegraphics[width=0.224\textwidth]{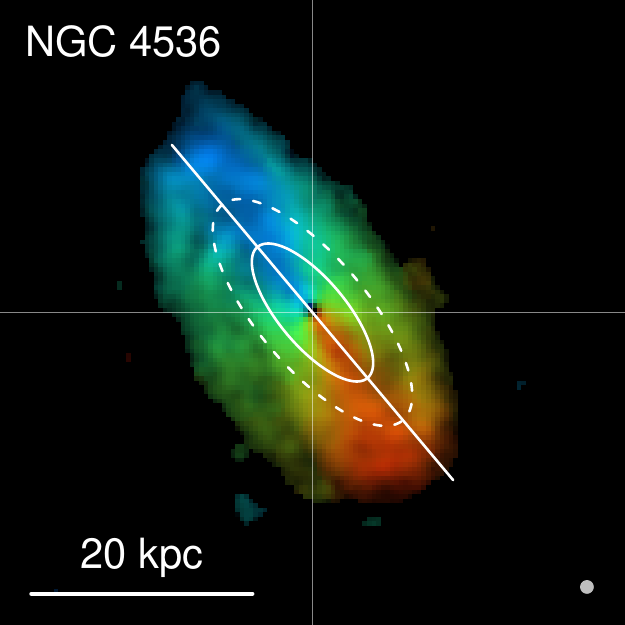}
\includegraphics[width=0.224\textwidth]{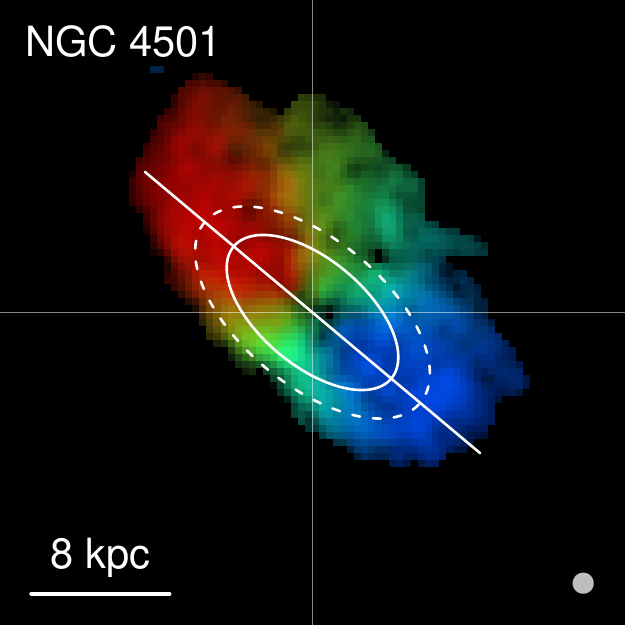}
\includegraphics[width=0.224\textwidth]{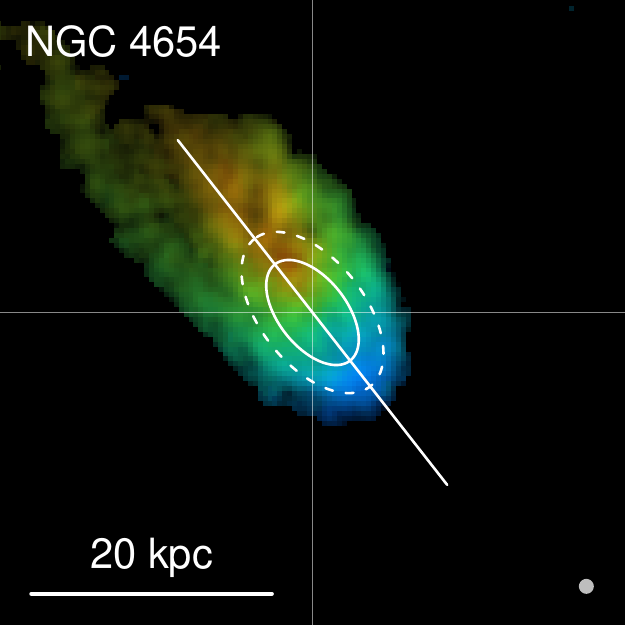}
\includegraphics[width=0.224\textwidth]{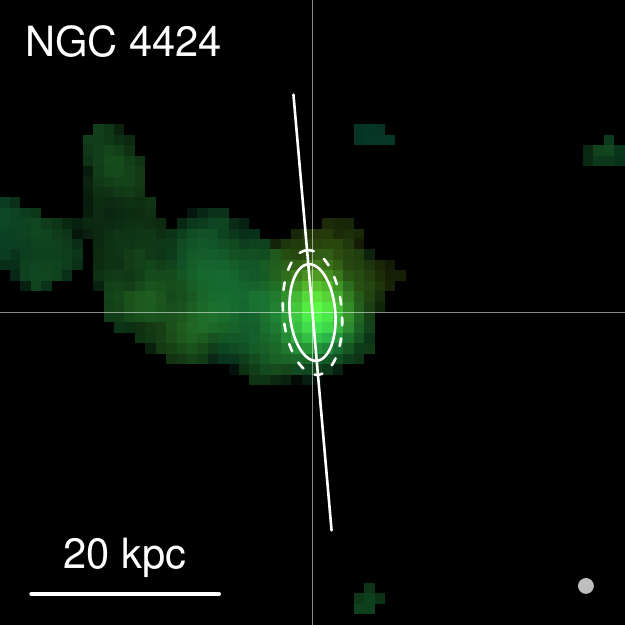}
\caption{\HI velocity-intensity maps of four galaxies in the VIVA sample. Stripping becomes stronger from the left to the right. Intensity represents the \HI surface density (moment~0 map) as shown by the colorbar. The grey circles show the approximate resolution (FWHM) of $15''$. The orientation and aspect ratio of the ellipses show the position angle and inclination of the galaxies (from \citealp{Chung2009}). The sizes of these ellipses are such that they contain half of the $r$-band emission (solid) and \HI (dashed).}
\label{fig:examples}
\end{figure*}

\subsection{Measurements}

In order to measure $j$ of the stellar and cold gas components, we assume that the galaxies rotate in a thin disk and that the stars co-rotate around the galactic centre at the same velocity as the gas, hence neglecting so-called asymmetric drift due to the different dispersion of stars and gas. This approximation is consistent with spectroscopic observations in large late-type galaxies, but tends to over-estimate the stellar rotation by up to 20\% in more dispersion-rich systems \citep{Cortese2016}. This error is comparable to other measurement uncertainties in $j$, e.g. due to inclination uncertainties (see the end of this section).

The sAM of stars ($j_*$) and \HI ($j_{\rm HI}$) can be evaluated from the 2D kinematic data, some examples of which are shown in Figure~\ref{fig:examples}. Explicitly, the $j$ values are computed as
\be\label{eq:jsum}
	j_{\rm phase} = \frac{\sum_k m_k r_k v_k}{\sum_k m_k},
\ee
where the sum goes over the pixels $k$ in the images and each pixel has a mass $m_k$, a galacto-centric radius $r_k$ and a circular velocity $v_k$. These three quantities are evaluated as follows.
\begin{itemize}
	\item The mass map $\{m_k\}$ of \HI is taken as the 21cm moment 0 map. For stars, the mass map is approximated by an exponential disk model, $m\propto\exp(-r/r_d)$, where the scale radius $r_d$ is computed as $r_d=R_e/1.678$ with $R_e$ being the $r$-band effective radius from \citet{Cortese2012}. Exponentials offer a good approximation of $j_*$ and allow us to extrapolate the optical data into the noise-dominated parts of the images \citep[e.g.\ OG14;][]{Romanowsky2012}. In galaxies with a significant stellar bulge (i.e.\ the Sa-Sb types in the sample, such as NGC4606), $R_e/1.678$ underestimates the scale $r_d$ of the exponential profile at large radii, leading to an underestimation of $j_*$. This is partially compensated by the overestimation of $j_*$ caused by neglecting asymmetric drift.
	\item The radii $\{r_k\}$ in the plane of the galaxy are obtained from the radii $s_k$ in the plane of the sky using the standard deprojection equation $r=F(x,y;\alpha,i)s$ with a function $F$ that depends on the $(x,y)$-position in the image, as well as on the position angle $\alpha$ and inclination $i$ of the galaxy. The explicit expression can be found in Eq.~(B3) of OG14.
	\item The circular velocities $\{v_k\}$ in the plane of the galaxy are computed from the line-of-sight velocities $v_z$ (1st moment of \HI line) using the deprojection equation $v=C(x,y;\alpha,i)v_z$, given in Eq.~(B4) of OG14.
\end{itemize}

The \HI images of VIVA are large enough and deep enough (typical \HI column density of 3--$5\cdot10^{19}\rm~cm^{-2}$) for the values of $j_{\rm HI}$ to be converged within a few percents (based on the detailed convergence study in OG14). However, in some galaxies the optical disk extends beyond the \HI disk, meaning that no velocity data is available in their outer parts. In these cases, the moment 1 map of \HI is extrapolated beyond the observations assuming a flat rotation curve with a velocity fixed at the 90\% quantile of the observed pixels. The relative increase of $j_*$ due to this extrapolation varies between 0\% and $\sim100\%$ with a mean of $25\%$. We caution, that this method bears some risk that the rotation curve has not yet reached the flat part, which would lead to an underestimation of $j_*$ and thus of $q$ and $\dfq$. However, this effect must be small since the maximum \HI velocity of our galaxy sample is consistent with the local baryonic Tully-Fisher relation \citep{McGaugh2015}.

When deprojecting the data, care must be applied to divergencies: $F$ diverges for edge-on galaxies ($i=90^\circ$), whereas $C$ diverges for face-on galaxies ($i=0^\circ$), as well as on the minor axis for galaxies of any inclination. Measurement errors of pixels close to a divergency can lead to large uncertainties in $j$. To avoid this problem, the sums in Eq.~(\ref{eq:jsum}) are only taken over pixels where $F<3$ and $C<3$. In other words, the observed radii and velocities are never multiplied by more than a factor 3 in the deprojection procedure. In axially symmetric galaxies, this rejection of pixels has no systematic effect on $j$, since the numerator and denominator in Eq.~(\ref{eq:jsum}) are reduced by the same factor. Since the minimum of $\left|C\right|$ is $\sin^{-1}i$ (along the major axis), the requirement that $\left|C\right|<3$ implies that only galaxies with $\sin^{-1}i<3$, i.e.\ $i>20^\circ$ can be used. All galaxies in our sample satisfy this condition, since the minimum inclination is $30^\circ$.

In summary, we have measurements of the stellar $j_*=J_*/M_*$ and atomic $j_{\rm HI}=J_{\rm HI}/M_{\rm HI}$. As we do not have sufficient data for good measurements of molecular sAM values, $j_{\rm H_2}$, we assume that these values are identical to $j_*$. This is justified by the approximate congruence between molecular and stellar material in nearby star-forming galaxies \citep{Walter2008}. For heavily stripped galaxies, where both the \HI and H$_2$ disks have been significantly truncated, $j_{\rm H_2}$ might be closer to $j_{\rm HI}$. However, a test of these heavily stripped galaxies ($\dfq>0.5$) shows that their sAM changes only by about a percent if approximating $j_{\rm H_2}\approx j_{\rm HI}$ instead of $j_{\rm H_2}\approx j_*$. For the one galaxy without CO data, the H$_2$ mass is neglected. The error made in doing so is at most a few percent, based on the other 37 galaxies with CO data.

The baryonic sAM $j$ can then be computed as the mass-weighted mean sAM of the components,
\be\label{eq:jthreecomponent}
	j=\frac{J}{M}=\frac{(M_*+1.35M_{\rm H2})j_*+1.35M_{\rm HI}j_{\rm HI}}{M_*+1.35(M_{\rm HI}+M_{\rm H2})}.
\ee
where the factor 1.35 ensures that Helium is accounted for in the atomic and molecular component.

Measurement uncertainties of $j$ are computed through linear propagation of inclination uncertainties of $10^\circ$ and assuming an additional extrapolation error of $10\%$, roughly the uncertainty in the maximum rotation velocity used for this extrapolation.

In environmentally perturbed and stripped systems, the assumption of a flat disk at constant inclination and position angle potentially introduces significant systematic errors in $j$, which are hard to estimate. However, the \HI mass fraction in stripped regions, is generally very small and often exaggerated in non-linear luminosity scales, such as in Figure~\ref{fig:examples}. In fact, even in one of the most heavily stripped galaxies NGC~4424, only 35\% of the \HI resides in the non-symmetric tail and of the stellar mass fraction in this tail is significantly smaller. Neglecting this material only changes the resulting value of $q\propto j/M$ by a few percent.

All sAM measurements with statistical uncertainties are listed in Table~\ref{table:VIVAgalaxies}. The table also lists the resulting values of $q$, computed using a fixed \HI velocity dispersion of $\sigma=10$~km/s for all galaxies. This assumption (justified in Section~\ref{HIdeficiency}) was made since the velocity resolution (10~km/s) of VIVA is insufficient for a direct measurement. The table also shows the atomic gas fractions $\fatm$ and \HI deficiencies, calculated using Eqs.~(\ref{equ:fatm}) and (\ref{equ:delfq}), respectively. Where available molecular masses are included.

\subsection{Results of the Virgo sample}\label{resultsvirgo}

\begin{figure}
\includegraphics[width=1.0\columnwidth]{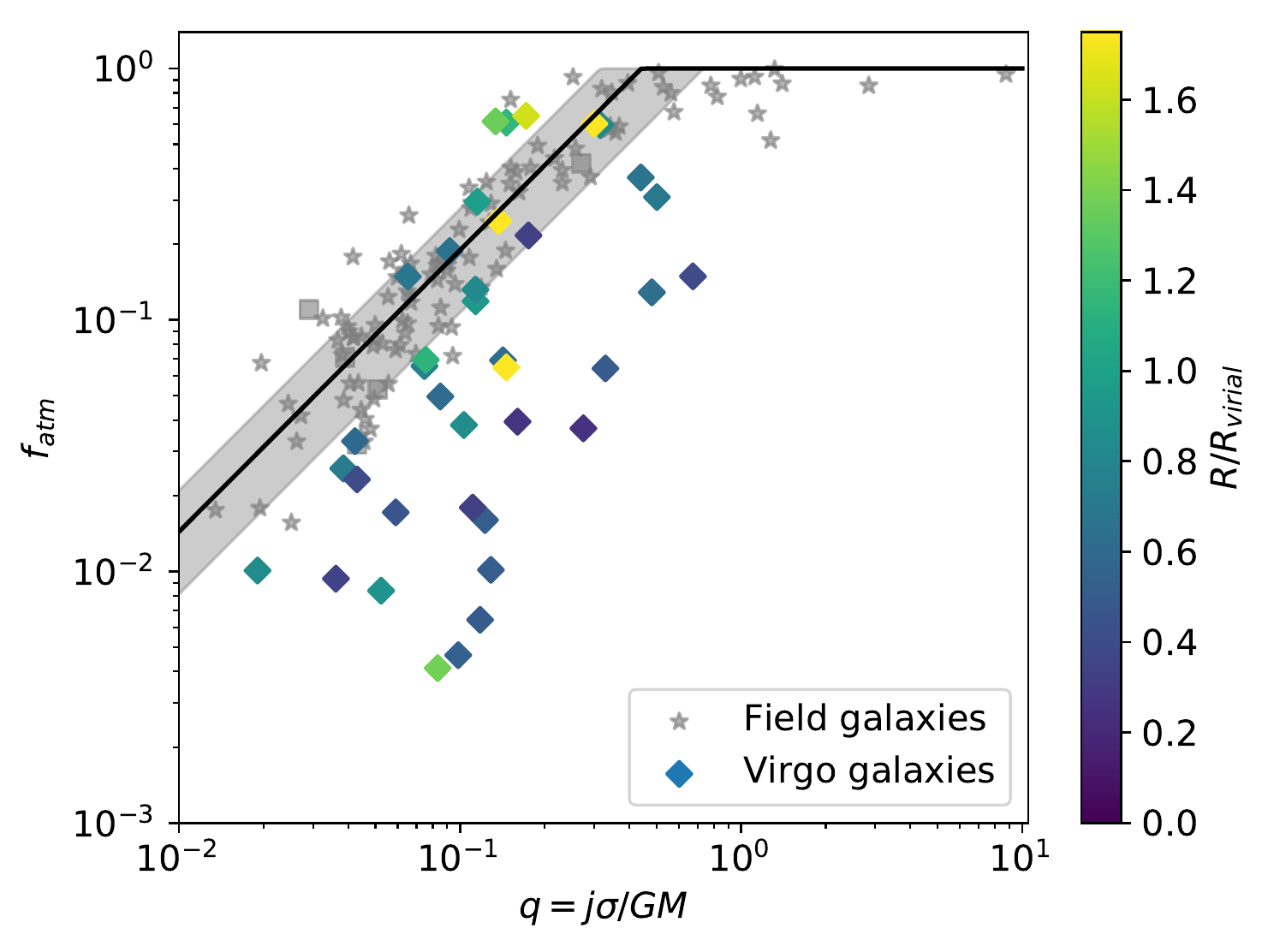}
\caption{Atomic gas fraction versus parameter $q$. Diamonds are the Virgo galaxies from the VIVA survey. Gray points are the field galaxies from Figure~\ref{fig:definitionfq}. The color bar represents the ratio between projected distance to the giant elliptical galaxy M87 at the centre of Virgo A and the virial radius of Virgo A \citep[1.55 Mpc,][]{Ferrarese2012}.}
\label{fig:fieldVIVA}
\end{figure}

\begin{figure}
	\includegraphics[width=1.0\columnwidth]{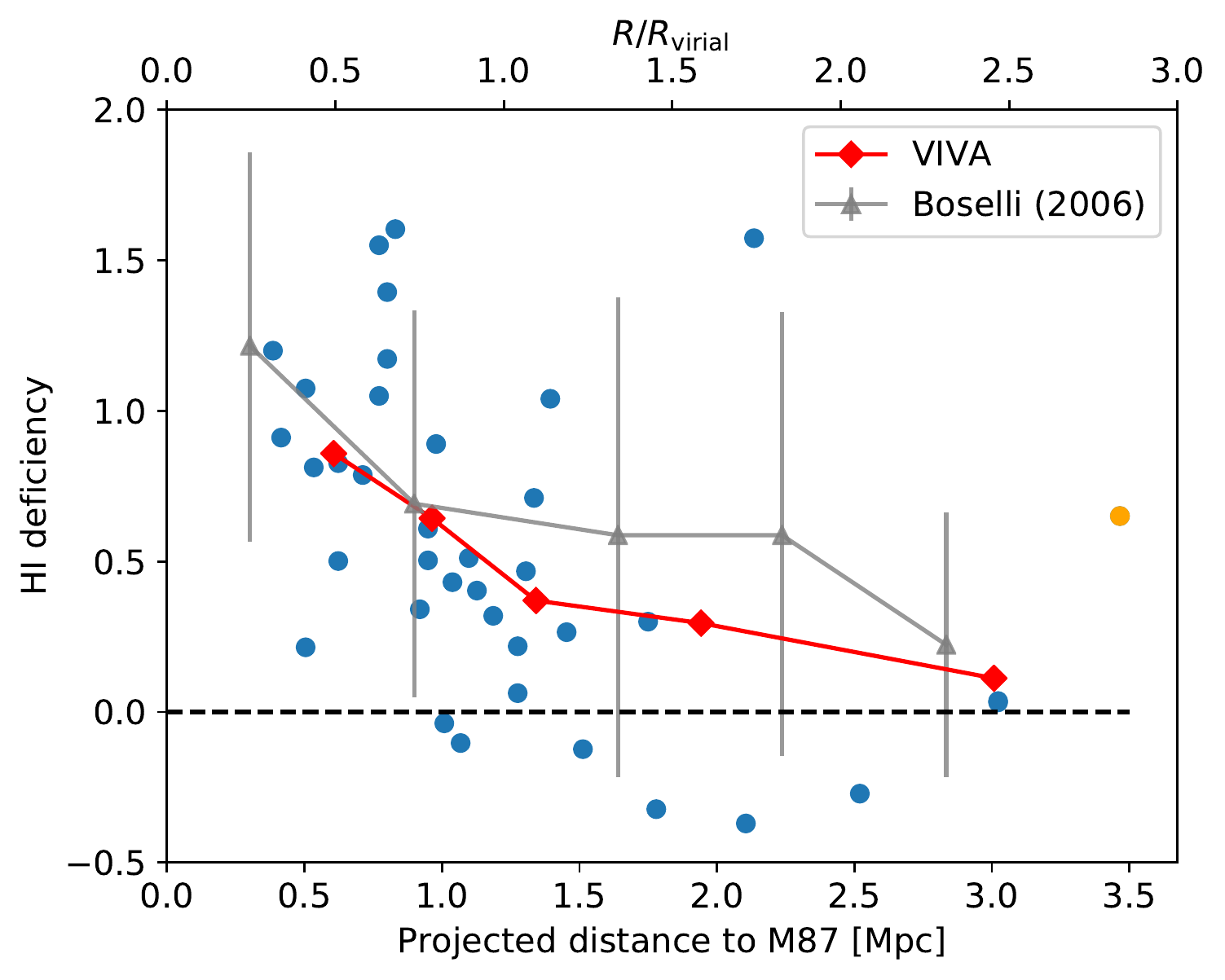}
	\caption{\HI deficiency as a function of projected distance from galaxies to the cluster centre M87. The upper axis shows the distances normalized to the virial radius of 1.55 Mpc. The red diamonds show the mean value of $\dfq$ in VIVA, binned by projected distance. The two right-most bins (distance $>1.6$ Mpc) are slightly wider to ensure that each bin contains at least three objects. The grey points and their uncertainty of 1$\sigma$ are from Figure 2 in \citet{Boselli2006}, presenting the average \HI deficiency in each bin from a larger, optically controlled sample. Their \HI deficiencies are measured by Eq.~\ref{equ:deltafD} ($\dfD$, detailed discussion in Sec~\ref{sss:dfd}). The orange point is NGC 4772, who is likely to have experienced a minor merger recently.}
	\label{fig:delfqdist}
\end{figure}

\begin{table*}
	\centering
	\caption{Key properties of the galaxies in our subsample of VIVA. Columns are as follows. (1) Galaxy names; (2) stellar masses from \citet{Cortese2012}, with a typical uncertainty $\sim$0.15 dex; (3) total \HI mass from \citet{Chung2009}; (4) \HI mass uncertainty; (5) H$_2$ mass from \citet{Boselli2014b}; (6) H$_2$ mass uncertainty; (7) half-mass radii in $r$-band from \citet{Cortese2012}; (8) half-mass radii of \HI (determined from the VIVA  moment 0 maps); (9) stellar sAM; (10) sAM of \HI; (11) typical uncertainty of $j_*$ and $j_{\rm HI}$; (12) parameter $q$; (13) neutral atomic gas fraction; (14) \HI deficiency $\Delta f_q$. In the units, $\rm lg$ stands for the base-10 logarithm.}
	\label{table:VIVAgalaxies}
	\begin{tabularx}{\textwidth}{@{\extracolsep{\fill}}cccccccccccccc}
		\hline
		ID & $M_*$ & $M_{\rm HI}$ & $\Delta M_{\rm HI}$ & $M_{\rm H_2}$ & $\Delta M_{\rm H_2}$ &$R_{\rm e}^{\rm opt}$ & $R_{\rm e}^{\rm HI}$ & $j_*$ & $j_{\rm HI}$ & $\Delta j$ & $q$ & $f_{\rm atm}$ & $\Delta f_q$\\
		-- & $\!\!\!{\rm lg}(\msun)\!\!\!$ & $\!\!\!{\rm lg}(\msun)\!\!\!$ & $\!\!\!{\rm lg}(\msun)\!\!\!$ &$\!\!\!{\rm lg}(\msun)\!\!\!$ &$\!\!\!{\rm lg}(\msun)\!\!\!$ & kpc & kpc & $\!\!\!{\rm lg}(\rm kpc\,km/s)\!\!\!$ & $\!\!\!{\rm lg}(\rm kpc\,km/s)\!\!\!$ & $\!\!\!{\rm lg}(\rm kpc\,km/s)\!\!\!$ & -- & -- & --\\
		(1) & (2) & (3) & (4) & (5) & (6) & (7) & (8) & (9) & (10) & (11) & (12) & (13) & (14)\\
		\hline
		IC 3392 & 9.77 & 7.69 & 0.34 & 8.62 & 0.19 & 2.60 & 1.60 & 2.56 & 2.20 & 0.05 & 0.129 & 0.010 & 1.39 \\
NGC 4192 & 10.65 & 9.68 & 0.04 & 9.39 & 0.20 & 9.70 & 16.50 & 3.41 & 3.52 & 0.04 & 0.113 & 0.119 & 0.27 \\
NGC 4216 & 11.00 & 9.30 & 0.09 & 9.21 & 0.20 & 6.30 & 13.60 & 3.22 & 3.66 & 0.04 & 0.038 & 0.026 & 0.40 \\
NGC 4222 & 9.31 & 8.86 & 0.10 & 8.06 & 0.19 & 3.80 & 7.40 & 2.68 & 3.06 & 0.04 & 0.501 & 0.308 & 0.51 \\
NGC 4254 & 10.39 & 9.70 & 0.04 & 10.02 & 0.05 & 5.10 & 9.90 & 3.06 & 3.30 & 0.21 & 0.065 & 0.149 & -0.10 \\
NGC 4298 & 10.10 & 8.75 & 0.08 & 9.16 & 0.18 & 3.90 & 4.10 & 2.75 & 2.71 & 0.07 & 0.085 & 0.050 & 0.50 \\
NGC 4302 & 10.44 & 9.22 & 0.07 & 9.29 & 0.18 & 7.90 & 10.30 & 3.29 & 3.38 & 0.04 & 0.142 & 0.069 & 0.61 \\
NGC 4321 & 10.71 & 9.51 & 0.02 & 9.91 & 0.05 & 8.10 & 10.70 & 3.33 & 3.34 & 0.18 & 0.075 & 0.066 & 0.32 \\
NGC 4330 & 9.52 & 8.70 & 0.10 & 8.61 & 0.19 & 6.00 & 4.90 & 3.13 & 3.05 & 0.04 & 0.674 & 0.149 & 0.83 \\
NGC 4351 & 9.17 & 8.53 & 0.06 & 8.11 & -- & 2.60 & 2.90 & 2.21 & 2.17 & 0.09 & 0.175 & 0.217 & 0.21 \\
NGC 4380 & 10.06 & 8.16 & 0.19 & 8.84 & 0.19 & 5.90 & 5.60 & 2.94 & 2.80 & 0.07 & 0.160 & 0.015 & 1.32 \\
NGC 4383 & 9.42 & 9.52 & 0.05 & 8.48 & 0.09 & 1.40 & 9.90 & 1.94 & 3.16 & 0.06 & 0.277 & 0.595 & -0.00 \\
NGC 4388 & 10.14 & 8.62 & 0.26 & 8.78 & 0.20 & 5.00 & 4.50 & 3.26 & 3.04 & 0.04 & 0.275 & 0.037 & 1.20 \\
NGC 4396 & 9.25 & 8.99 & 0.09 & 8.55 & 0.18 & 4.50 & 5.80 & 2.75 & 2.94 & 0.04 & 0.439 & 0.369 & 0.43 \\
NGC 4402 & 10.04 & 8.62 & 0.18 & 9.31 & 0.05 & 4.70 & 4.10 & 3.00 & 2.77 & 0.04 & 0.160 & 0.039 & 0.91 \\
NGC 4419 & 10.24 & 7.82 & 0.62 & 9.11 & 0.05 & 2.60 & 2.50 & 2.91 & 3.01 & 0.05 & 0.098 & 0.005 & 1.60 \\
NGC 4424 & 9.91 & 8.34 & 0.07 & 8.86 & 0.18 & 5.20 & 6.70 & 2.35 & 2.01 & 0.06 & 0.054 & 0.031 & 0.48 \\
NGC 4450 & 10.70 & 8.51 & 0.08 & 9.07 & 0.20 & 4.70 & 5.80 & 3.07 & 3.07 & 0.11 & 0.052 & 0.008 & 1.04 \\
NGC 4457 & 10.43 & 8.34 & 0.11 & 9.19 & -- & 2.00 & 2.50 & 2.38 & 2.38 & 0.18 & 0.019 & 0.010 & 0.47 \\
NGC 4501 & 10.98 & 9.27 & 0.06 & 9.88 & 0.05 & 6.00 & 8.20 & 3.30 & 3.36 & 0.06 & 0.043 & 0.023 & 0.50 \\
NGC 4522 & 9.38 & 8.58 & 0.13 & 8.90 & 0.18 & 4.10 & 4.10 & 2.93 & 2.81 & 0.04 & 0.481 & 0.129 & 0.89 \\
NGC 4532 & 9.21 & 9.34 & 0.03 & 8.30 & 0.18 & 2.70 & 6.20 & 2.27 & 2.58 & 0.05 & 0.146 & 0.610 & -0.32 \\
NGC 4535 & 10.45 & 9.57 & 0.02 & 9.55 & 0.05 & 8.60 & 13.20 & 3.25 & 3.37 & 0.10 & 0.113 & 0.132 & 0.22 \\
NGC 4536 & 10.26 & 9.73 & 0.02 & 9.45 & 0.05 & 7.80 & 12.80 & 3.20 & 3.33 & 0.05 & 0.137 & 0.248 & 0.04 \\
NGC 4548 & 10.74 & 8.86 & 0.03 & 8.88 & 0.05 & 6.10 & 8.70 & 3.16 & 3.19 & 0.14 & 0.059 & 0.017 & 0.79 \\
NGC 4561 & 8.99 & 9.20 & 0.03 & 8.57 & -- & 2.40 & 5.40 & 1.83 & 2.54 & 0.18 & 0.149 & 0.591 & -0.30 \\
NGC 4569 & 10.66 & 8.85 & 0.10 & 9.69 & 0.05 & 8.70 & 4.90 & 3.41 & 3.00 & 0.05 & 0.111 & 0.018 & 1.07 \\
NGC 4579 & 10.94 & 8.80 & 0.12 & 9.36 & 0.05 & 4.80 & 6.60 & 3.15 & 3.27 & 0.14 & 0.036 & 0.009 & 0.81 \\
NGC 4580 & 9.99 & 7.50 & 0.34 & 8.60 & 0.18 & 2.50 & 1.20 & 2.57 & 2.32 & 0.13 & 0.083 & 0.004 & 1.57 \\
NGC 4606 & 9.77 & 7.45 & 0.23 & -- & -- & 3.00 & 0.80 & 2.48 & 1.83 & 0.06 & 0.118 & 0.006 & 1.55 \\
NGC 4607 & 9.60 & 8.39 & 0.16 & 8.80 & 0.19 & 3.50 & 3.30 & 2.87 & 2.76 & 0.04 & 0.329 & 0.064 & 1.05 \\
NGC 4651 & 10.13 & 9.66 & 0.03 & 8.96 & 0.21 & 3.20 & 8.20 & 2.87 & 3.24 & 0.08 & 0.115 & 0.295 & -0.12 \\
NGC 4654 & 10.14 & 9.52 & 0.03 & 9.61 & 0.05 & 5.10 & 7.80 & 2.95 & 3.06 & 0.07 & 0.092 & 0.188 & -0.04 \\
NGC 4689 & 10.19 & 8.73 & 0.05 & 9.31 & 0.05 & 5.10 & 4.50 & 2.93 & 2.79 & 0.15 & 0.103 & 0.038 & 0.71 \\
NGC 4698 & 10.52 & 9.27 & 0.03 & 8.65 & -- & 4.30 & 16.90 & 3.00 & 3.55 & 0.05 & 0.076 & 0.069 & 0.30 \\
NGC 4713 & 9.22 & 9.51 & 0.03 & 8.72 & 0.18 & 2.60 & 8.70 & 2.40 & 2.80 & 0.08 & 0.172 & 0.648 & -0.27 \\
NGC 4772 & 10.25 & 8.97 & 0.06 & 8.48 & -- & 4.40 & 6.20 & 3.07 & 3.28 & 0.06 & 0.146 & 0.065 & 0.65 \\
NGC 4808 & 9.49 & 9.61 & 0.03 & 8.59 & 0.19 & 2.40 & 12.40 & 2.42 & 3.25 & 0.05 & 0.300 & 0.603 & 0.03 \\

		\hline
	\end{tabularx}
\end{table*}

The diamonds in Figure \ref{fig:fieldVIVA} show the 38 Virgo galaxies of our sample in the $(q,f_{\rm atm})$-plane. As expected, most of these galaxies lie below the analytical relation, in the \HI undersaturated region. Almost all our Virgo galaxies, except for most of  those right at the edge of the cluster (green--yellow colours in the figure), lie significantly ($>0.2$~dex) below the analytical relation, whereas almost of all \textit{field} galaxies of the reference sample (grey stars) lie on the analytical relation. This suggests that the $(q,f_{\rm atm})$-plane is a useful diagnostic for the presence of environmental effects that do not significantly affect $q$. Indeed, in stellar mass-dominated galactic disks, $q$ only has a weak dependence on \HI, which vanishes if $j_*=j_{\rm HI}$. Thus, even if \HI is heavily stripped, $q$ only changes slightly. For instance, removing \emph{all} the \HI from the Milky Way would only decrease its $q$-value by $\sim5\%$.

The offset of the galaxies from the analytical $q$--$\fatm$ relation of \HI saturated systems is quantified by $\dfq$. Figure~\ref{fig:delfqdist} shows this theoretically motivated `\HI deficiency' $\dfq$ as a function of the projected distance of the galaxies to the central cluster galaxy M87. As naively expected and well-established \citep[e.g.][]{Davies1973,Giovanelli1985,Chung2009}, there is a trend for the \HI deficiency to increase with the proximity to the cluster centre. However, this qualitative statement is subject to a list of caveats:
\begin{itemize}
\item The decrease of $\dfq$ with increasing radius is slightly stronger than observed in optically complete samples (gray points in Figure~\ref{fig:delfqdist}; \citealt{Boselli2006}), but the statistical significance of this tension is marginal because of the small number of objects.
\item Some galaxies might already have had a peri-centre passage, where they might have been stripped of their \HI content, even if now they are situated at large radii.
\item Other environmental effects than stripping could have affected selected galaxies. For instance, NGC 4772 (the orange point in Figure \ref{fig:delfqdist}) is likely to have experienced a minor merger recently \citep{Haynes2000}.

\end{itemize}
Numerical simulations, such as the one discussed in Section~\ref{EAGLE} can resolve these caveats, as they provide access to three-dimensional geometries and a look back in time.

\subsection{Comparison between different \HI deficiencies}\label{ss:otherestimators}

Let us now compare the new \HI deficiency estimator $\dfq$ to two more familiar empirical estimators, mainly uses in simulation-based studies (Section~\ref{sss:dfm}) and observational environmental studies (Section \ref{sss:dfd}), respectively.

\subsubsection{\HI deficiency relative to stellar mass}\label{sss:dfm}

A simple way of defining a \HI deficiency, often used for simulations \citep[e.g.][]{Crain2017,Stevens2018b}, is to measure the offset of a galaxy from the mean stellar mass-\HI mass relation, or, equivalently the baryon mass-\HI fraction relation, $M$--$f_{\rm atm}$. Formally, this definition of the \HI deficiency can be written as
\begin{eqnarray}
\dfM={\rm log_{10}}\left[0.5\left(\frac{M}{10^9\msun}\right)^{-0.37}\right]-{\rm log_{10}}\left(f_{\rm atm}\right),
\label{equ:deltafM}
\end{eqnarray}
where the first term on the right-hand-side denotes the mean value of $f_{\rm atm}$ at baryonic (stellar+cold gas) mass $M$ in a volume-complete sample of galaxies, given by O16 and consistent with the optically-selected GASS sample \citet{Catinella2010}.

The advantage of this definition is its simplicity, only requiring global stellar mass (light) and \HI-mass measurements. However, this simplicity comes at the cost of several disadvantages. Firstly, the definition of $\dfM$ is purely phenomenological, providing no insight into the physical causes of \HI-rich and \HI-poor galaxies. Secondly, the $M$--$f_{\rm atm}$ distribution exhibits a large intrinsic scatter ($\sim 0.5$~dex, see O16 Figure 3), even for spiral galaxies that show no evidence of environmental effects, hence making $\dfM$ a poor diagnostic of such effects. Thirdly,  \HI-selected samples are biased towards \HI-rich galaxies in the $M$--$f_{\rm atm}$ space and therefore biased negatively in $\dfM$. At least in field galaxies, this inherent bias towards \HI-rich galaxies does not (or at most weakly) bias $\dfq$, because the \HI-rich systems are indeed the ones with high spin and thus high $q$, as shown in the Appendix Figure~\ref{fig:Mfatm_j}. However, we make the comparison here for reference.

 Figure~\ref{fig:delfqfM}(a) shows the comparison of $\dfM$ and $\dfq$ in our sample of 38 Virgo galaxies. From this figure, $\dfM$-$\dfq$ is close to one-to-one relation despite a small offset ($\sim 0.1$~dex) due to aforementioned selection effects. In this case, $\dfM$ can be used to identify \HI deficient galaxies.

Cosmological simulations can bypass concerns of selection bias, since they are limited by mass and volume rather than sensitivity. Also, the mean $M$--$f_{\rm atm}$ relation (or $M_*$--$M_{\rm HI}$ relation) is well-defined. For instance, one can select all central galaxies (as opposed to satellites) to determine the median value of $M_{\rm HI}$ as a function of stellar mass and then compute the deficiency $\Delta f_{\rm M}$ of the satellites relative to this relation. Such an approach has been presented, for instance, by \citet{Crain2017} for the EAGLE simulations, \citet{Stevens2018a} for the Illustris-TNG simulations and \citet{Lagos2011,Lagos2018b} for semi-analytic models. 

\subsubsection{\HI deficiency relative to optical size}\label{sss:dfd}

Traditional definitions of \HI deficiencies, used in observational environmental studies, compare the \HI mass of individual galaxies to that of field galaxies of the same optical size and (sometimes) same morphological type \citep{Haynes1984a}. Formally this can be written as
\be
\Delta f_{\rm D}={\rm log_{10}}M_{\rm HI}^{\rm ref}-{\rm log_{10}}M_{\rm HI}^{\rm obs},
\label{equ:deltafD}
\ee
\noindent where $M_{\rm HI}^{\rm ref}$ is the expected \HI mass of field galaxies and $M_{\rm HI}^{\rm obs}$ is the observed value of an individual object. The variation of $M_{\rm HI}^{\rm ref}$ with morphological type is often modelled as using a power law approximation
\be
{\rm log_{10}}h^2M_{\rm HI}^{\rm ref}={\rm A}+{\rm B}{\rm log_{10}}(h^2D^2)
\ee
\noindent where $D$ is optical diameter (e.g.~at the $25\rm~mag/arcsec^2$ isophote), $h\approx0.7$ is the Hubble parameter and $A$ and $B$ are numerical constants. These constants are normally calibrated separately to each morphological Hubble type, such as shown in Table~3 of \citet{Boselli2009}.

\begin{figure*}
\includegraphics[width=1.0\textwidth]{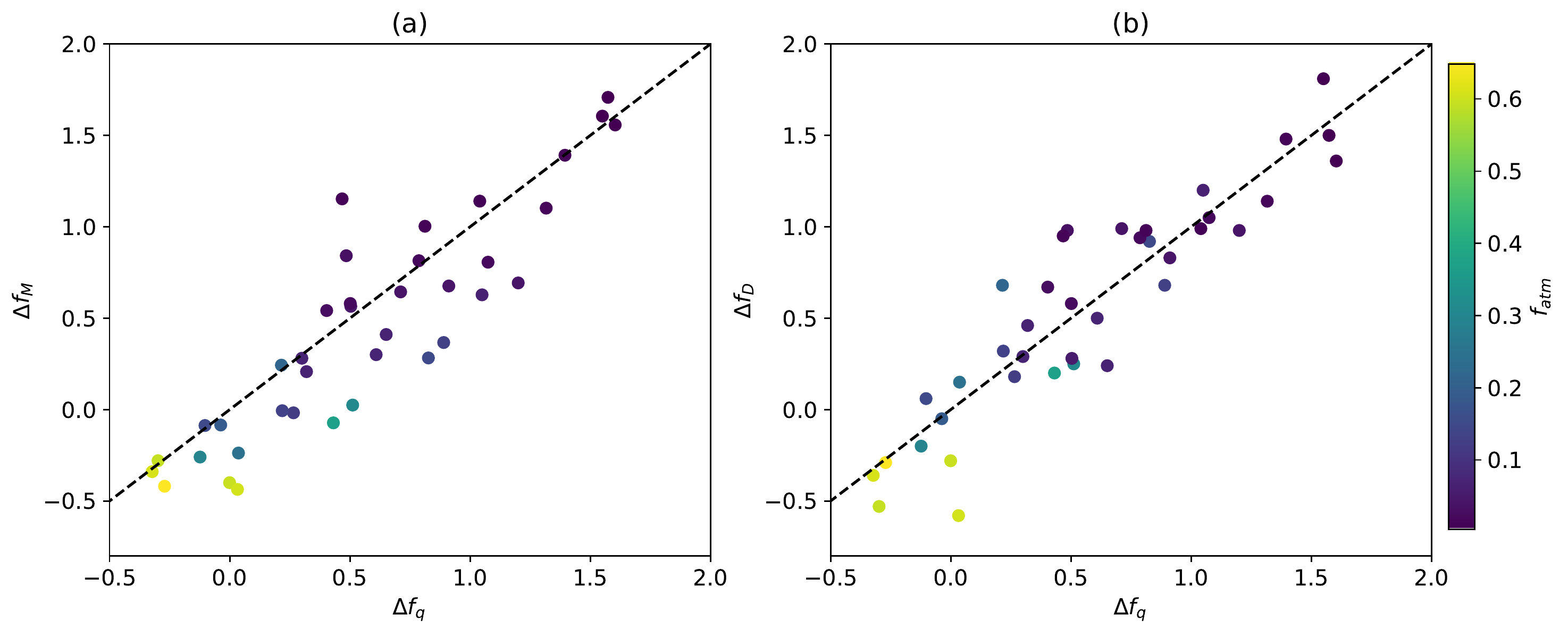}
\caption{Comparison between the different \HI deficiency estimators $\dfD$, $\dfM$ and $\dfq$. The colour bar shows the atomic gas fraction. Dashed lines are the one-to-one relations.}
\label{fig:delfqfM}
\end{figure*}

 Indeed Figure \ref{fig:delfqfM}(b) shows that $\dfD$ and $\dfq$ fall nearly on a one-to-one relation. This similarity between $\dfD$ and $\dfq$ can be understood from the fact that the galactic size is determined by the sAM \citep{Fall1980,Mo1998}. However, in very high \HI-rich galaxies, $\dfD$ and $\dfq$ will diverge. The value of $\dfD$ can even be negative, while $\dfq\sim 0$, because (1) the optical diameter is much smaller than the baryonic diameter, and (2) $\dfq$ is based on the gas fraction, who has maximum value 1.

In environmental observational studies of \HI, $\dfD$ is widely used. However, in cosmological simulations, the definition and measurement of morphological types is somewhat cumbersome and thus $\dfD$ is not normally used.

In summary, the estimator $\dfD$ often has similar values to $\dfq$, but the latter has the advantages of  (1) being easily useable with observations and simulations, as well as (2) having a more straightforward physical interpretation in terms of the \HI saturation point.

\section{Insights from the EAGLE simulations}
\label{EAGLE}

In this section, we investigate how the cluster galaxies in a hydrodynamics cosmological simulation evolve in the $q-f_{\rm atm}$ plane. The advantage of such a simulation is that we can trace the history of the galaxies and explore the evolution of $f_{\rm atm}$ through cosmic time. We use the EAGLE simulation \citep{Schaye2014,Crain2015,McAlpine2016}, which successfully reproduce several statistical gas properties of a volume-complete sample, such as the total neutral gas (\HI+H$_2$)-stellar mass relation \citep{Bahe2016}, the scaling relation between H$_2$ mass and stellar mass, star formation rate and stellar surface density of galaxies \citep{Lagos2015}, and reasonably well the \HI-stellar mass relation of galaxies \citep{Crain2017}. Particularly relevant to our work are the results of \citet{Marasco2016}, who showed that EAGLE reproduces the observed decrease in the \HI-to-stellar mass ratio with increasing parent halo mass (up to $10^{14.75}\msun$). We here use the largest EAGLE simulation box with a volume of $(100\,\rm Mpc)^3$, which contains 10 galaxy clusters of halo masses $>10^{14}\,\rm M_{\odot}$.

\subsection{EAGLE simulation}

The EAGLE simulation suite (details in \citealt{Schaye2014}, \citealt{Crain2015} and \citealt{McAlpine2016}) consists of a large number of cosmological hydrodynamic simulations with different resolutions, cosmological volumes, run with a modified version of the parallel N-body smoothed particle hydrodynamics (SPH) code GADGET-3. The simulation follows the formation and evolution of galaxies and supermassive black holes in a standard $\Lambda$ cold dark matter universe, and uses several subgrid physics modules including (1) star formation \citep{Schaye2008}, (2) stellar evolution and chemical enrichment \citep{Wiersma2009a}, (3) radiative cooling and photoheating \citep{Wiersma2009b}, (4) stellar feedback \citep{Dalla2012}, and (5) black hole growth and active galactic nucleus (AGN) feedback \citep{Rosas2015}. Table \ref{table:EAGLE} provides the parameters of the Ref-L100N1504 simulation used in this paper. 

\begin{table}
	\centering
	\caption{Basic properties of the EAGLE run Ref-L100N1504. Here, cMpc and ckpc refer to comoving Mpc and kpc, respectively, while pkpc refers to physical kpc.}
	\label{table:EAGLE}
	\begin{tabularx}{\columnwidth}{@{\extracolsep{\fill}}lccr}
		\hline
		 & Property & Units & Value\\
		\hline
		(1) & Comoving box side $L$ & cMpc & $100$\\
		(2) & Number of particles &  & $2 \times 1504^3$\\
		(3) & Gas particle mass & $\rm M_{\odot}$ & $1.81\times 10^6$\\
        (4) & DM particle mass & $\rm M_{\odot}$ & $9.7\times 10^6$\\
        (5) & Gravitational softening length & ckpc & $2.66$\\
        (6) & Max. gravitational softening length & pkpc & $0.7$\\
		\hline
	\end{tabularx}
\end{table}

\subsubsection{Calculation of the \HI fraction}

The temperature and density of the gas in EAGLE are calculated directly as part of the hydrodynamic simulation, accounting for radiative cooling and feedback processes using subgrid models. A global temperature floor, $T_{\rm eos}(\rho)$, is imposed, corresponding to a polytropic equation of state, $P\propto \rho^{\gamma_{\rm eos}}_{\rm g}$, where $\gamma_{\rm eos}=4/3$. This equation is normalised
to give a temperature $T_{\rm eos}=8\times 10^3$~K at $n_{\rm H}=10^{-1}\,\rm cm^{-3}$, which is typical of the warm interstellar medium \citep[e.g.][]{Richings2014}.

Hence, the ionised, atomic and molecular gas phases of the interstellar medium are not separated while running the simulation (and thus not fully consistent with the instantaneous star formation rates). Therefore, the amount of atomic material was calculated in post-processing by \citet{Lagos2015}. They first computed the neutral (atomic+molecular) gas fraction of each particle following the prescription of \citet{Rahmati2013}, which was calibrated to cosmological simulations coupled with full radiative transfer calculations. \citet{Rahmati2013} presented fitting functions to calculate the neutral fraction on a particle-by-particle basis from the gas temperature and density, and the total ionization rate (photoionization plus collisional ionization). Given the neutral gas fraction of a particle, \citet{Lagos2015} applied the \citet{GK11} model to split this gas further into its atomic and molecular phases. This model assumes that the dust-to-gas ratio and the radiation field are the driving processes setting the \HI/H$_{2}$ ratio. It further assumes that the dust catalyses the formation of H$_{2}$, while the UV radiation destroys the dust, preventing the formation of H$_{2}$. \citet{Lagos2015} assume the dust-to-mass ratio to scale with the local gas metallicity and the radiation field to scale with the local surface density of the star formation rate. 

Given this way of computing \HI masses in EAGLE, we then compute the atomic gas fractions $\fatm$ of each cluster galaxy using the definition of Eq.~(\ref{equ:fatm}), as in the observations. The mass limit of the EAGLE simulation implies a \HI mass limit close to $10^7\msun$. Galaxies with less \HI will be represented by zero \HI mass. We account for this effect in our statistical analysis of Section~\ref{ss:quenching}.

\subsubsection{Calculation of the $q$ parameter}
\label{sss:parameterq}

The parameter $q$ remains as defined in Eq.~(\ref{equ:q}). We assume $\sigma=10$~km/s for the \HI  and the global $j$ is calculated via:
\be
j=\frac{M_{*}j_{*}+1.35M_{{\rm gas}}j_{{\rm gas}}}{M_{*}+1.35\,M_{{\rm gas}}},
\label{equ:j_EAGLE}
\ee
\noindent where $M_{*}$ and $j_{*}$ are the mass and sAM of stars, and $M_{\rm gas}$ and $j_{\rm{gas}}$ are the mass and the sAM of the neutral hydrogen (\HI\!+H$_2$). This equation slightly differs from Eq.~(\ref{eq:jthreecomponent}), because the neutral gas phases have been combined into a mean $j_{\rm{gas}}$ in \citet{Lagos2017a}, but note that this does not affect the global baryonic $j$. \citeauthor{Lagos2017a} computed $j_*$ and $j_{\rm gas}$ by summing over all relevant particles $k$,
\be
j_{\rm phase}=\left|\frac{\Sigma_k m_k \textbf{r}_k\times \textbf{v}_k}{\Sigma_k m_k}\right|,
\ee
 \noindent where $m_k$ are the particle masses and $\textbf{r}_k$ and $\textbf{v}_k$ are the position and velocity vectors, relative to the centre-of-mass. To estimate $j_{*}$, they used star particles only, while the estimation of $j\rm{_{gas}}$ made use of all gas particles that have a neutral gas fraction $>0$. \citeauthor{Lagos2017a} showed that the stellar $M_*$--$j_*$ relation in EAGLE galaxies of different morphologies has good agreement with measurements in the local universe by \citet{Romanowsky2012,Obreschkow2014,Cortese2016}.

The resolution limit of the simulation constrains the galaxy selection. \citet{Schaye2014} suggested that the results are consistent between the simulations being used here and higher resolution, smaller volume simulations for $M_*\ge 10^9\,M_{\odot}$. \citet{Lagos2017a} showed that the estimation of $j_*$ is converged for $M_*\ge 10^{9.5}\,\rm M_{\odot}$ (see their Appendix A). We analyze  galaxies with $M_*\ge 10^{9.5}\,\rm M_{\odot}$ in this paper. 

\subsubsection{EAGLE clusters}

In EAGLE, haloes are identified using a friend-of-friends algorithm \citep[FoF,][]{Davis1985}. There are 10  clusters with $M_{\rm halo}\geqslant10^{14}\,\rm M_{\odot}$ and 205 smaller clusters (or `massive groups') with $10^{13}\,\rm M_{\odot}\leqslant M_{\rm halo}<10^{14}\,\rm M_{\odot}$ at $z=0$. In this paper, we use the 10 massive clusters for comparing $\dfq$ between EAGLE and VIVA (Section~\ref{ss:eagle-viva}), while we also include the smaller clusters in the following time-evolution analysis (Section \ref{sec:trace}) for increased statistical power.

We measure the distance of satellite galaxies from the centre of their cluster and compare the results with the observation. The coordinates of each galaxy are available from the EAGLE public database \citep{McAlpine2016}. We use the centre of potential of the FoF group as the cluster centre and measure the 3D distance to the satellites in units of $R_{200}$ (of the particular cluster). The shapes of the clusters are irregular because they are defined by as FoF groups. We select the galaxies at $z=0$ that lie within less than $1.75R_{200}$ from the cluster centres, regardless of whether these galaxies belong to the cluster, to do our analysis and comparison.

\subsection{\HI fraction in EAGLE cluster galaxies at $z=0$}\label{ss:eagle-viva}

We now compare the VIVA data against a comparable sample of EAGLE cluster galaxies. For consistency between the two data set, we only consider the 10 most massive EAGLE clusters ($M_{\rm halo}>10^{14}\,\rm M_{\odot}$), of which we select the satellite galaxies at radii $<1.75R_{200}$ and with $M_*>10^{9.5}{\rm M_{\odot}}$, $M_{\rm HI}>10^{7.45}\msun$ and $q>0.01$. The $M_*$ cut is, in fact, necessary because of the resolution of the simulation (see Section~\ref{sss:parameterq}), but it happens to be an almost perfect match to the VIVA sample, which has just four objects slightly below $10^{9.5}\msun$ ($10^{9.22 \pm 0.15}\msun$ being the smallest value). The $M_{\rm HI}$ cut matches the minimum value in the VIVA subsample (see Table~\ref{table:VIVAgalaxies}). Likewise, the selection $q>0.01$ is approximately consistent with the $q$-range in VIVA. Furthermore, this criterion effectively corresponds to a morphology selection to disk dominated galaxies (as constant values of $q\propto j/M$ approximately correspond to different morphologies \citealp{Obreschkow2014}). Overall, this selection results in a sample of 200 galaxies across all 10 clusters. Note that EAGLE used a FoF algorithm to determine whether a galaxy belongs the cluster or not. Of the 200 galaxies, 117 physically belong to a cluster in this sense, while the remaining 83 are close to the cluster, but do not belong to the FoF.

Before using the selected sample of EAGLE cluster galaxies to study \HI deficiencies $\dfq$, we first check if similarly selected field galaxies are indeed \HI saturated in the sense of Eq.~(\ref{equ:fq}). To this end, we consider the simulated galaxies with halo masses $10^{11}{\rm M_{\odot}}<M_{\rm halo}\leqslant 10^{12}{\rm M_{\odot}}$ and exclude those who have satellites and have been moving into a cluster/massive group in the past. Apart from this criterion, we apply exactly the same selection as for the EAGLE cluster galaxies, that is $M_*>10^{9.5}{\rm M_{\odot}}$, $M_{\rm HI}>10^{7.45}\msun$ and $q>0.01$. This results in a sample of 2100 field galaxies. Field galaxies suffer less from environmental effects and are thus expected to be \HI saturated. 

\begin{figure}
	\includegraphics[width=1.0\columnwidth]{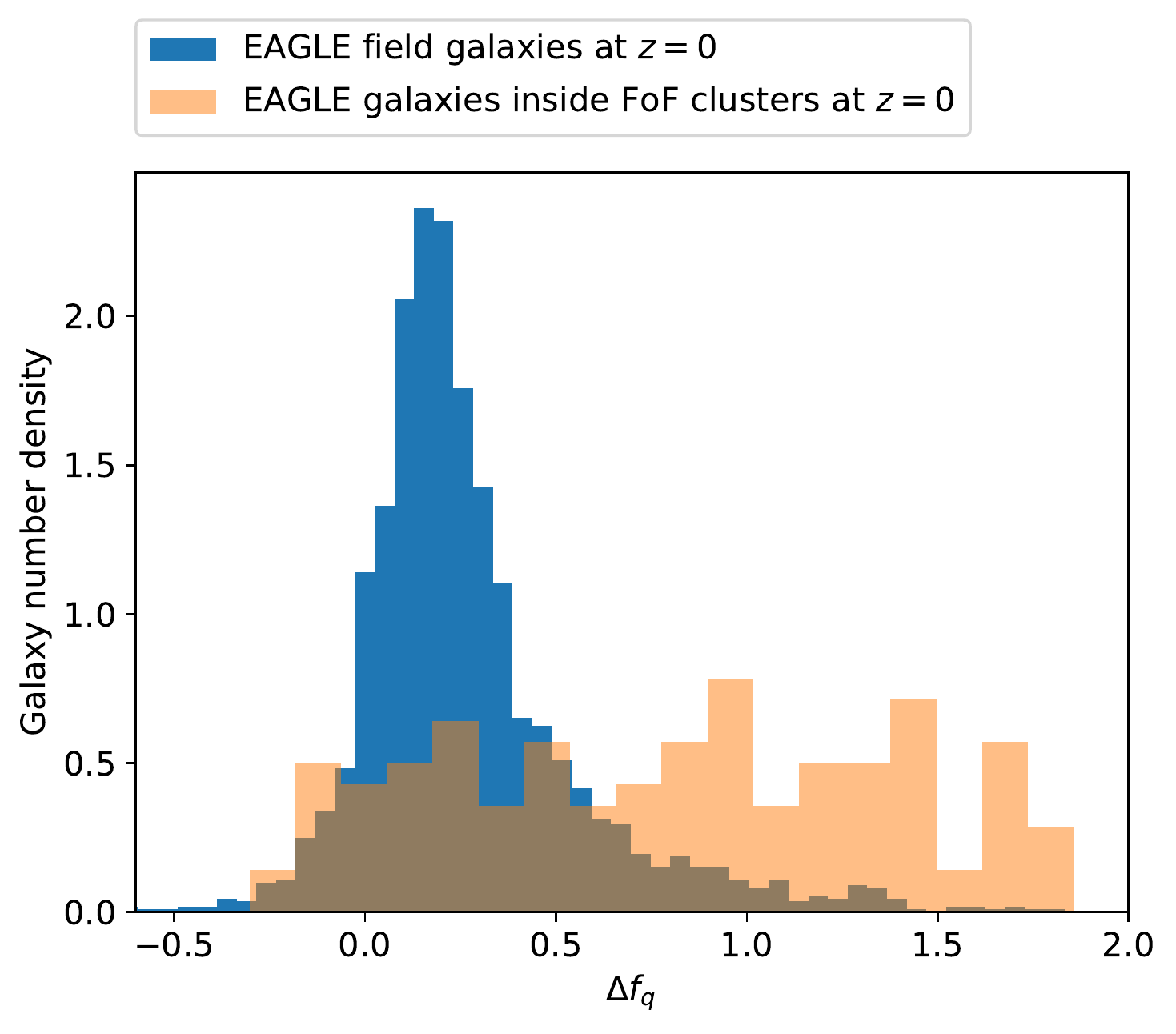}
	\caption{Number density of EAGLE field and cluster galaxies as a function of $\Delta f_q$. The median and standard deviation of the field galaxies are 0.20 and 0.29, respectively. The median and standard deviation of the cluster galaxies (in the FoF) are 0.85 and 0.58, respectively. The cluster galaxies show a much more significant positive tail into the stripped regime. The $p$-value of the Kolmogorov-Smirnov test for these two distributions is $10^{-5}$, comfirming that the two distributions are different.}
	\label{fig:EAGLEclusterIsodistri}
\end{figure}

Figure~\ref{fig:EAGLEclusterIsodistri} shows the $\Delta f_q$-histograms of EAGLE field and cluster galaxies. The simulated field galaxies exhibit a peak near $\dfq \sim 0.2$, while cluster galaxies are scattered to significantly larger values of $\dfq$, demonstrating that the simulated cluster galaxies are indeed more frequently \HI deficient than the field galaxies. The fact that the mean of field galaxies is $\dfq \sim 0.2$ rather than $\dfq\sim0$ is likely caused by a slight underestimation of the \HI mass in EAGLE (for galaxies in this halo mass range), also seen in an the offset of $M_{\rm HI}-M_*$ relative to observations \citep[see Figure~7 in][]{Crain2017}. In principle, one could  introduce an ad hoc correction of $0.2\rm dex$ for the simulated $\dfq$-values at $z=0$. However, since this correction is small compared to the typical values of stripped galaxies and systematic uncertainties in the determination of $\dfq$, we decided not to apply such a correction.

\begin{figure}
\includegraphics[width=1.0\columnwidth]{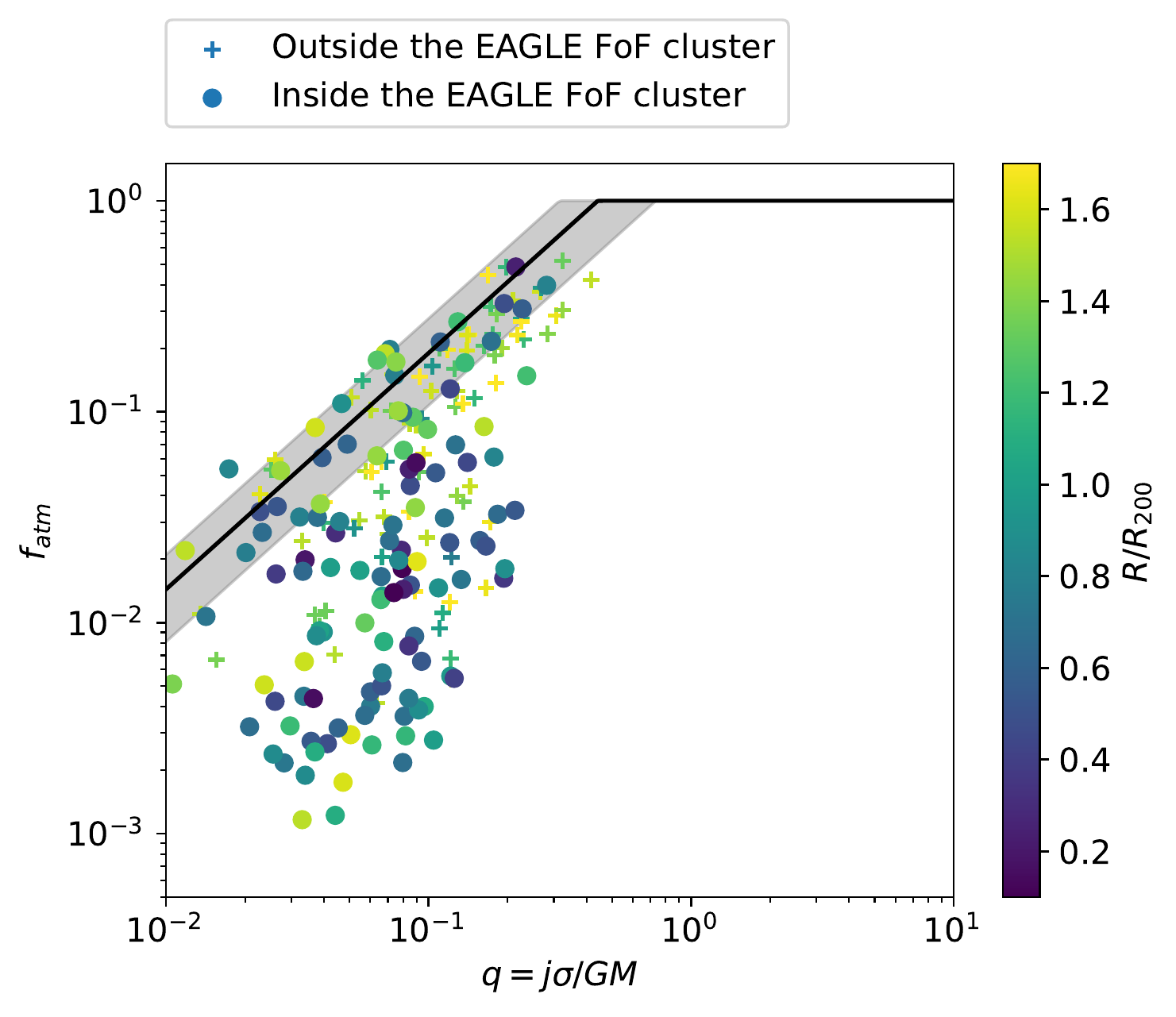}
\caption{($q,f_{\rm atm}$)-distribution of the EAGLE satellite galaxies in clusters with $M_{\rm halo}>10^{14}\msun$, selected as described in Section \ref{ss:eagle-viva}. The circles are identified inside the (FoF) clusters while the crosses are outside the (FoF) clusters. The solid line and shading show the $q-f_{\rm atm}$ relation for field galaxies as in Figure~\ref{fig:definitionfq}. The colour bar shows the 3D distance from the satellite galaxies to the cluster centre.}
\label{fig:EAGLEfield}
\end{figure}

Figure \ref{fig:EAGLEfield} shows the EAGLE cluster galaxies in the $q-f_{\rm atm}$ plane (at $z=0$). Most of the simulated satellite galaxies lie in the \HI undersaturated region, similarly to the observational data (Figure~\ref{fig:fieldVIVA}). An advantage of the simulation is that we can investigate the causes of  \HI deficiencies. Following \citet{Marasco2016}, the dominant cause of \HI deficiencies in EAGLE clusters more massive than $10^{14}\,\rm M_{\odot}$ is ram pressure stripping, with a significant secondary contribution from high-speed satellite-satellite encounters, causing dynamical heating of the cold gas. Tidal stripping plays a negligible role in our EAGLE satellites, but becomes important in group-scale haloes ($<10^{13}\,\rm M_{\odot}$).

\begin{figure}
\includegraphics[width=1.0\columnwidth]{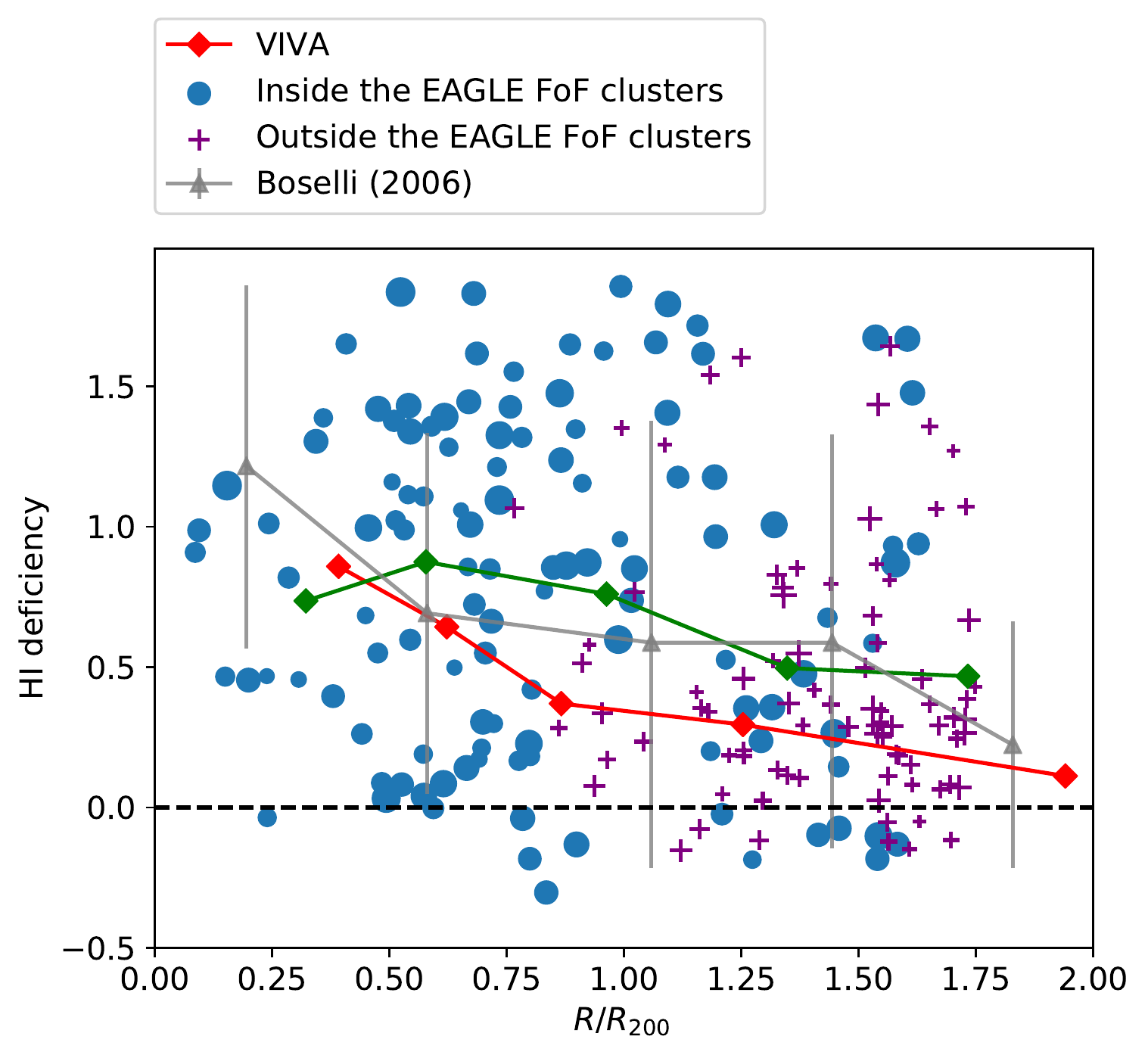}
\caption{$\dfq$ as a function of the 3D distance to the centre of potential of EAGLE galaxies. The blue points represent the galaxies inside the friends-of-friends (FoF) group of the clusters. In turn, the purple crosses denote galaxies outside the FoF groups. The bigger the size of the symbols, the larger the baryonic mass of a galaxy. The green diamonds and line represent the mean value of all the selected EAGLE galaxies in different radius bins. As in Figure \ref{fig:delfqdist}, the red line is the $\Delta f_q$-distance relation from VIVA and the grey points are from the optical sample of \citet{Boselli2006}. All data are normalized to the virial radius 1.55 Mpc.}
\label{fig:EAGLEdistfq}
\end{figure}

In Figure \ref{fig:EAGLEdistfq}, we show the $\dfq$-distance relation in EAGLE compared to VIVA and samples from \citet{Boselli2006}. We find that the tendency of galaxies to become more \HI deficient as they get closer to the cluster centre is qualitatively similar in EAGLE and VIVA. Quantitatively the trend is more pronounced in VIVA because, as we mentioned in Section~\ref{resultsvirgo}, VIVA lacks statistical power because of the limited number of galaxies at a radii $\gtrsim2$~Mpc. In line with this explanation, the EAGLE data is more consistent with the \HI deficiency trend of an complete optically selected sample (binned grey points in Figure \ref{fig:EAGLEdistfq}).

The simulated data in Figure \ref{fig:EAGLEdistfq} shows very large scatter at all distances from the cluster centre. This scatter is probably attributed to similar effects as in empirical data (see bullet points in Section \ref{resultsvirgo}), that is the clusters are not fully relaxed, galaxies at identical radii have very different histories, some galaxies have mutual interactions that affect their \HI, etc. Furthermore, galaxies at the same distance from the cluster centre may reside in different intra-cluster medium (ICM) densities, because large clusters are generally \emph{not} spherically symmetric. In EAGLE this is clear from the fact that many galaxies outside the spherical radius $R_{200}$ lie still within the FoF group (blue dots at $R>R_{200}$), while other galaxies inside this spherical radius lie outside the FoF group (purple crosses at $R<R_{200}$). Likewise, real clusters are often aspherical \citep{Jauzac2016,Stroe2015}.

\begin{figure}
\includegraphics[width=0.95\columnwidth]{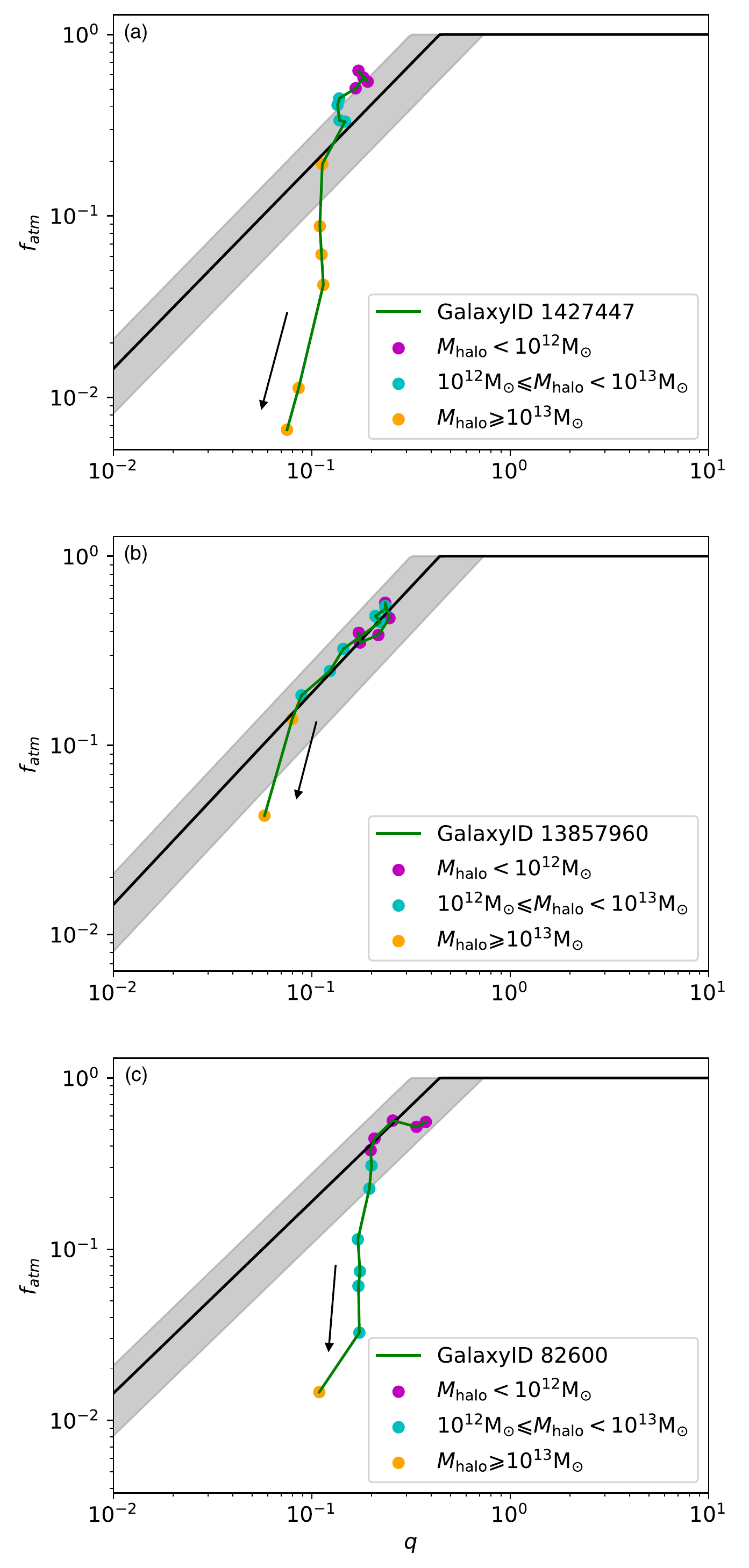}
\caption{Three tracks of EAGLE galaxies in the $q-f_{\rm atm}$ plane from $z\approx2$ to $z=0$. Arrows show the direction of time. The colours of the points represent the mass of the parent halo. The top and middle panels show the stripped satellite galaxies in our sample (heavily stripped+less stripped, 92\% in total), while bottom panel shows rarer cases of pre-processing ($\sim$8\%) in haloes with $M_{\rm halo}<10^{13}\msun$. }
\label{fig:track}
\end{figure}

\subsection{Tracing the \HI evolution history of cluster galaxies}
\label{sec:trace}

Part of the reason for the large scatter in the simulated distance--$\dfq$ relation (Figure \ref{fig:EAGLEdistfq}) is that satellite galaxies with similar distances to their cluster centres have different histories. To understand the variety of these histories, we traced the satellite galaxies identified at $z=0$ back in time to redshift $z=2$ or to the highest redshift where the galaxy's stellar mass lies still above the resolution limit ($10^{9}\msun$). This provides a maximum of 13 snapshots per galaxy. In order to trace the infall history, we only retain the galaxies that have lived in a smaller halo ($<10^{13}\msun$) before entering the cluster. Of our 117 cluster galaxies (in EAGLE FoFs $>10^{14}\msun$), 113 (97\%) satisfy this selection criterion. We refer to this sample of 113 galaxies with resolved histories as `Sample 1'. 

The time interval between snapshots is $\approx 0.5-1$~Gyr, depending on the snapshot \citep[see][]{McAlpine2016}. We use the merger trees from \citet{Qu2017}, which are available in the public database, to reconstruct the history of the satellite galaxies.  In addition, we record the halo mass, stellar mass, the \HI mass from \citet{Lagos2015} and the sAM from \citet{Lagos2017a} for each galaxy at different redshifts, from which we can reconstruct their $q$ and $f_{\rm atm}$ histories.

Sample 1 has been designed to have a very similar selection to the VIVA sample, while also allowing to track the galaxies back in time. However, with only 113 galaxies, this sample has limited statistical power. In order to increase the statistical power of the evolution analysis, we also consider a second sample (`sample 2'), in which the minimum limit for the final halo mass at $z=0$ has been reduced from $10^{14}\msun$ to $10^{13}\msun$.This change increases the sample size to 690, while likely maintaining ram pressure stripping as the most significant source of \HI deficiencies following the detailed analysis by \cite{Marasco2016}. Note that Sample 1 is a subset of Sample 2.

Before presenting a statistical analysis of Sample 1 and 2, let us illustrate the evolution of some individual galaxies in the $q-\fatm$ plane as they fall into a cluster. Figure~\ref{fig:track} shows three such evolutionary tracks. In most cases, stripping starts occuring in haloes $>10^{13}\msun$, as highlighted by the yellow colouring chosen for these haloes. We classify the evolutionary tracks into three classes, corresponding to the three panels in Figure~\ref{fig:track}. The first and second class (panels a and b) include all the galaxies that are not significantly \HI deficient ($\dfq<0.5$) before entering a halo more massive than $10^{13}\msun$. In EAGLE, 90\% (92\%) of the galaxies fall into this category in Sample 1 (Sample 2). The difference between the first and second class lies in the final \HI deficiency at $z=0$. The first class (panel a) contains all the strongly stripped cases ($\dfq>0.5$ at $z=0$); the second class (panel b) contains all the weakly stripped cases ($\dfq<0.5$ at $z=0$). In EAGLE, 52\% (49\%) fall into the first class; 38\% (43\%) fall into the second class. The most common reason for objects to fall into the second class is that they have not been long enough in a massive halo (e.g. only for 2 snapshots in Figure~\ref{fig:track}b). The remaining objects in the second class orbit the massive halo at large radii ($\gtrsim0.8R_{200}$), where stripping is less effective. Finally, there is a third class of galaxies (panel c), which are already significantly \HI deficient ($\dfq>0.5$) before entering a massive halo ($>10^{13}\msun$). These galaxies have been `pre-processed' in a smaller galaxy group (typically between $10^{12}$ and $10^{13}\msun$) prior to entering the massive halo. In EAGLE, 10\% (8\%) of the cluster galaxies fall into this category.

\begin{figure}
\includegraphics[width=1.0\columnwidth]{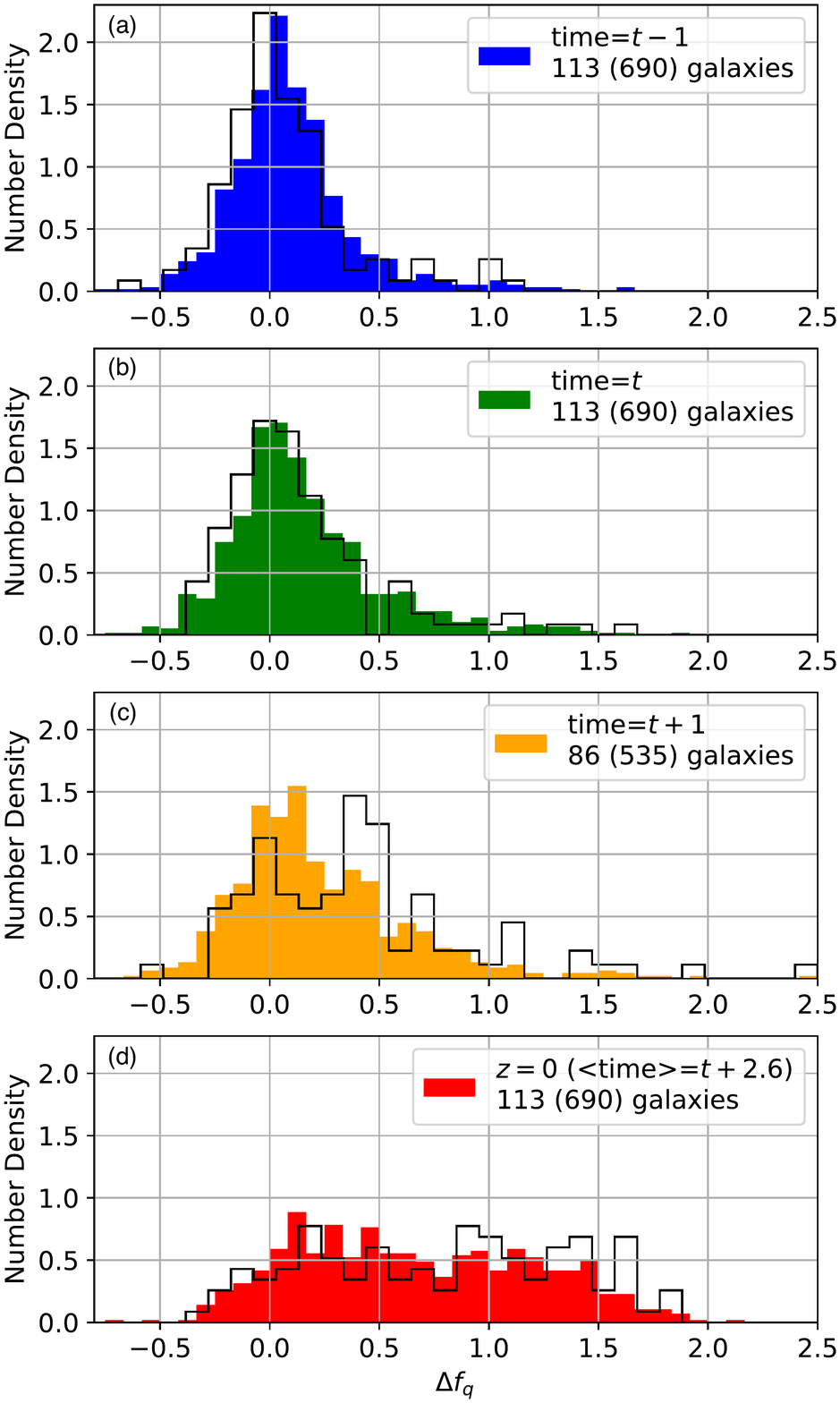}
\caption{$\dfq$ distributions of satellite galaxies in Sample 1 (black lines) and Sample 2 (filled histograms), defined in the 2nd paragraph of Section \ref{sec:trace}. The four panels represent four different snapshots, relative to the time the galaxies first enter their haloes with $M_{\rm halo}>10^{13}\msun$ (details in Section~\ref{sec:trace}). The number of galaxies are given in the legend for Sample 1 (Sample 2 in parentheses).}
\label{fig:hist_clu}
\end{figure}

A way of representing the statistical evolution of $\dfq$ of satellite galaxies from Sample 1 and Sample 2 is shown in Figure \ref{fig:hist_clu}. This figure displays the histogram of $\dfq$ in four different snapshots: immediately before entering a halo of $M_{\rm halo}>10^{13}\msun$, refered as the snapshot $t-1$, the first snapshot after falling into such a halo (refered as the snapshot $t$), one snapshot later (refered as the snapshot $t+1$) and at $z=0$ (which mixes various relative times t, t+1,t+2...). 

In Figure \ref{fig:hist_clu}(a), the galaxies from Samples 1 and 2 exhibit a quite symmetric distribution in $\dfq$ with a mean close to zero, showing that no significant \HI deficiencies are seen, on average. The only statistically significant exception are a few systems in the positive tail at $\dfq>0.5$, which correspond to pre-processed systems such as shown in Figure~\ref{fig:track}c.

At later times (panels (b)--(d)), the galaxies become increasingly \HI deficient. The sparse snapshots in EAGLE limit a precise determination of when exactly the galaxies enter a cluster and how quickly their \HI removal occurs. On average, the galaxies are 5--100-times  \HI deficient at $z=0$, i.e.~their \HI mass is 5--100 times smaller than the saturation mass (16- and 84-percentile range).

\begin{figure}
\includegraphics[width=1.0\columnwidth]{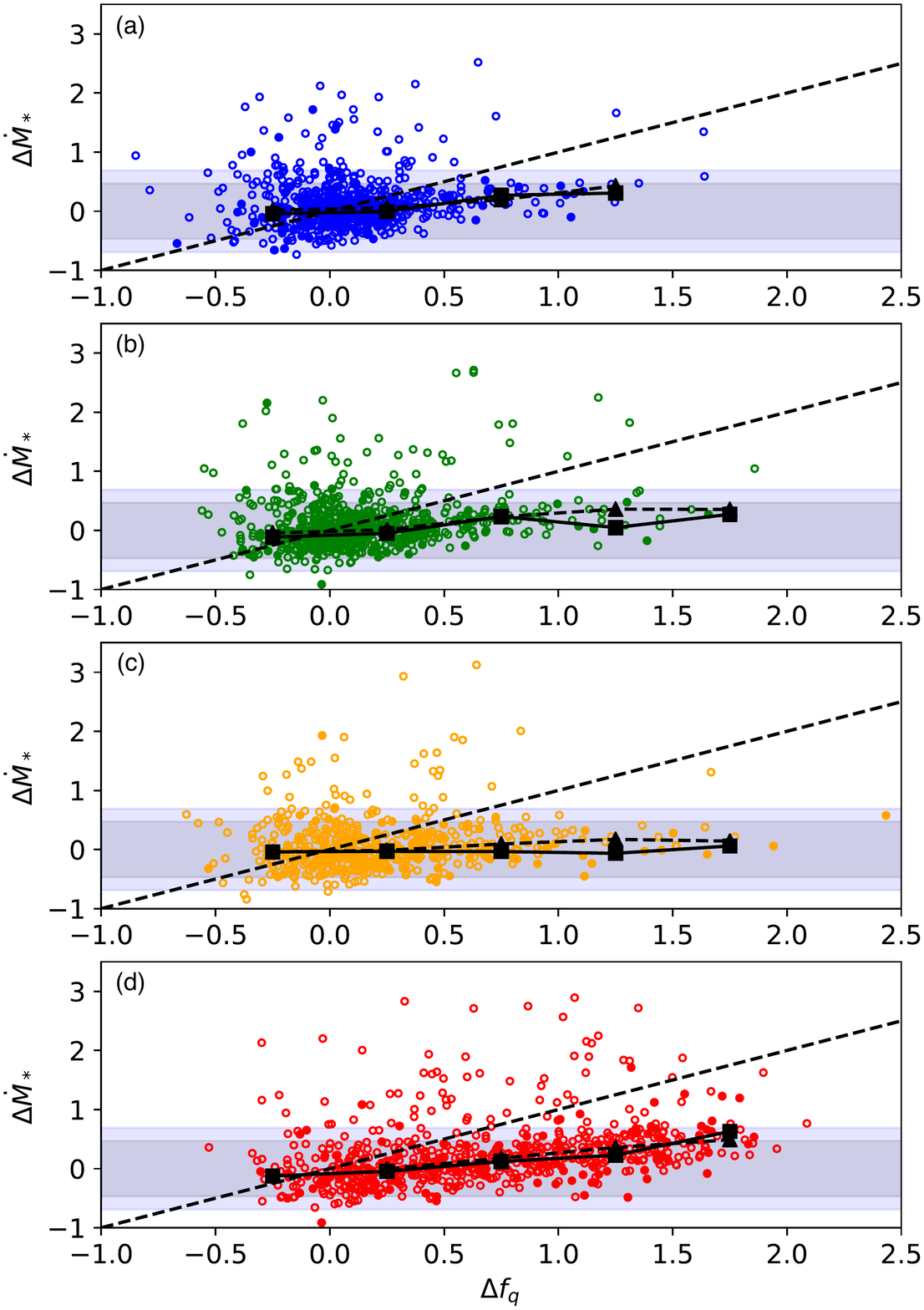}
\caption{$\Delta f_q-\Delta \dot{M}_*$ relation of the galaxy sample for the same snapshots (relative to entering the cluster) shown in Figure~\ref{fig:hist_clu}. The solid points are Sample 1 and the empty points are Sample 2. The solid line shows the median value of sample 1 and the dashed line for sample 2. Light and dark grey shaded regions represent 1/5 to 5 and 1/3 to 3 factors below and above the SF main sequence, respectively. These factors are commonly used in the literature to define the SF main sequence width \citep{Bethermin2015}. The black squares and lines show the mean value of $\Delta \dot{M}_*$ in bins of $\dfq$.}
\label{fig:Mdot}
\end{figure}

\subsection{Discussion of quenching process of EAGLE cluster galaxies}
\label{ss:quenching}

Let us finally discuss the link between the \HI deficiency estimator $\dfq$ and the suppression of star formation. Observations \citep[e.g.][]{Cortese2009,Boselli2014b} present a tight relation between the specific star formation rates and traditional \HI deficiency parameters, indicating that the removal of gas changes faster than the quenching of the star formation activity. In hydrodynamic simulations and semi-analytic models, galaxies with suppressed star formation (SF) are typical also very \HI poor or even have a complete absence of gas \citep[][ among ohters]{Lagos2014,Marasco2016,Crain2017,Lagos2018b,Stevens2018b}. Among the physical processes behind this relation are galaxy-galaxy encounters \citep{Marasco2016}, ram pressure stripping \citep{Stevens2018b}, strangulation \citep{Lagos2018b}.

Here, we define the amount of relative suppression of star formation as the offset of a galaxy from the `main sequence', i.e. the mean $M_*$--$\dot{M}_*$ relation at the considered redshift. The offset is measured in logarithmic units along the $\dot{M}_*$-axis,
\be
	\Delta \dot{M}_*={\log_{10}(\dot{M}_*^{\rm MS})}-{\log_{10}(\dot{M}_*^{\rm gal})},
\ee
where the reference value $\dot{M}_*^{\rm MS}$ is drawn from a redshift-dependent power-law fit to the $M_*$--$\dot{M}_*$ relation, for the subsample of galaxies with stellar masses between $10^9\msun$ and $10^{10.5}\msun$ and with specific star-formation rates (${\rm sSFR}=\dot{M}_*/M_*$) whose $\log_{10}({\rm sSFR}/[{\rm Gyr^{-1}}])\geqslant -2+0.5z$. This sSFR cut (for $z<2$) was proposed by \citet{Furlong2015} for the EAGLE simulations to exclude passive galaxies from the fit.
%

%
In other words, we fit the main sequence to EAGLE galaxies that are considered as star-forming at their redshift, as Eq.~\ref{equ:sSFRfit} shows, and we use this fit to estimate the distance to the main sequence for individual galaxies.

\be
\label{equ:sSFRfit}
\log_{10}\left(\frac{\rm sSFR_{\rm MS}}{\rm Gyr}\right)=-0.14\times \log_{10}\frac{M_*}{\msun}-0.47
\ee

The evolution of the star firmation main sequence as a function of redshift in EAGLE is similar to the observations \citep[see Fig. 7 in][]{Furlong2015}, following the power law $(z+1)^n$ with $n\sim 3.5$.

Figure \ref{fig:Mdot} presents the $\Delta f_q-\Delta \dot{M}_*$ relation for the same snapshots (relative to entering the cluster) as shown in Figure~\ref{fig:hist_clu}. The two samples present a similar trend.

When the gas is removed ($\dfq$ becomes positively skewed), galaxies suppress their SFR moderately ($\Delta \dot{M}_*$ becomes larger) compared to their decrease in \HI content, which may indicate the time delay between quenching and removal of \HI gas. An important part of this delay is that the \HI is stripped preferentially in the galactic outskirts, where the \HI depletion time can be much longer than on average. Thus, removing this outer \HI only has a weak effect star formation.

Interestingly, the $\dfq$--$\Delta \dot{M}_*$ is not a one-to-one relation, in that the relative amount of SF suppression is smaller than the \HI deficiency, irrespective of how long ago the suppression started. This finding is qualitatively consistent with observations in the range of $\dfq<1$ \citep[e.g.][]{Cortese2009} and explainable by the fact that stripping preferentially acts on the low-density outskirts of galactic disks, where the specific star formation rates of \HI are very low. Removing this \HI only implies a relatively small effect on star-formation. However, in heavily stripped systems ($\dfq>1$), the results of observations and our simulated sample are in tension. \citet{Cortese2009,Boselli2014c} showed that when $\dfD>1$ (which is similar to $\dfq$), few galaxies are still in the SF main sequence while most of them are quenched. In our comparison sample (Sample 1), however, many galaxies are still on the SF main sequence despite them having $\dfq>1$. It remains unclear how significant this tension is since the empirical evidence is subject to small number statistics and the EAGLE simulation is pushed to its mass resolution limit. Therefore, the $\dfq-\Delta \dot{M}_*$ relation is an interesting topic for future investigations.

A direct comparison of the slope in the $\dfq$--$\Delta \dot{M}_*$ between different models is not straight forward. For instance Figure~8 of \citet{Stevens2018b} shows that $\Delta \dot{M}_*$ and $\dfq$ covary, but a direct comparison to our study is complicated by different analysis techniques. A comparison of the $\dfq$--$\Delta \dot{M}_*$ between models and observations could offer an interesting avenue for future research, which might benefit from the use of $\dfq$ as a uniform definition of \HI deficiencies.

\section{Conclusions}
\label{conclusions}

In this paper, we analysed the \HI deficiencies of environmentally affected cluster galaxies in the theoretical framework of the $(q,\fatm)$-plane, where $q$ is the atomic stability parameter introduced by O16. Field disk galaxies lie on a tight relation in this plane, which matches the analytically predicted relation for \HI-saturated exponential disks. The offset $\dfq$ from this relation is thus a sensible definition of a galaxy's \HI deficiency accounting for its baryonic mass and sAM, i.e.\ its major dynamical properties.

By applying $\dfq$ to the galaxies in the VIVA survey, we confirmed the validity of $\dfq$ as a sensitive probe of environmental removal/suppression of \HI (e.g.~due to stripping). In doing so, the $q$-values of the VIVA galaxies were computed using angular momentum measurements from spatially resolved kinematic 21cm data. By calculating $\dfq$ for cluster galaxies in the EAGLE simulation, we confirmed that this simulation exhibits a qualitatively similar trend of \HI-deficiencies as a function of cluster-centric distance, although a more quantitative comparison would require more than one observed cluster. The simulation allowed us to trace stripped galaxies back in time and determine the moment when the \HI deficiencies first become detectable. Trough a statistical analysis of 690 galaxies in haloes more massive than $10^{13}\msun$, we found that the $\dfq$ distribution morphs significantly (towards positive values) as soon as the galaxies first enter the virial sphere of a halo more massive than $10^{13}\msun$. This is paralleled by a slightly delayed quenching in the star-formation activity.

Compared to standard non-kinematic empirical definitions of \HI deficiencies (such as the stellar mass based $\dfM$, or the optical size+morphology based $\dfD$), $\dfq$ exhibits a list of interesting advantages: (1) it has a direct physical interpretation in terms of the \HI saturation point; (2) In the case of disk-dominated systems, $\dfq$ does  not require any calibration to a reference sample of field galaxies, although we used such a reference sample to demonstrate that $\dfq\approx0$ in this case; (3) $\dfq$ can be  measured in observations and simulations in a like-to-like way. These advantages come at the prize of requiring kinematic data, ideally kinematic maps with $\lesssim1~\rm kpc$ spatial and $\lesssim10~\rm km/s$ velocity resolution in the rest-frame.

With the fast rise of integral field spectroscopy surveys (e.g. SAMI, HECTOR, MANGA) and radio interferometry surveys of \HI (e.g.~WALLABY and DINGO) on the Australian Square Kilometre Array Pathfinder (ASKAP) will move the field of survey astronomy towards large galaxy samples with sufficient data for accurate mass and angular momentum measurements and hence good determinations of $\dfq$. Data from these surveys, as well as unresolved \HI data from optically complete alternative surveys \citep[e.g.][]{Catinella2018} will enable much more systematic analyses of galaxies in the $(q,\fatm)$-plane. These studies will allow testing detailed theoretical predictions emerging from cosmological simulations and semi-analytic models. For instance, \citet{Stevens2018a}, using the semi-analytic model DARK SAGE, qualitatively predicted the different $q$--$\fatm$ distributions of galaxies, whose \HI was removed by entirely different mechanisms, such as ram pressure stripping, minor mergers and quasar-mode feedback.

\section*{Acknowledgements}

We thank for the useful discussion from Yannick Bahe, Pascal Elahi and Ivy Wong. We acknowledge the Virgo Consortium for making their simulation data available. The EAGLE simulations were performed using the DiRAC-2 facility at Durham, managed by the ICC, and the PRACE facility Curie based in France at TGCC, CEA, Bruy\'eresle-Chatel. DO and CL thank for support from the Australia Research Council Discovery Project 160102235. CL thanks funding from ASTRO 3D. LC and DO are recipients of Australian Research Council Future Fellowships (FT180100066 and FT190100083) funded by the Australian Government. Parts of this research were conducted by the Australian Research Council Centre of Excellence for All Sky Astrophysics in 3 Dimensions (ASTRO 3D), through project number CE170100013. We thank the anonymous referee for their insightful and constructive feedback.







\appendix

\section{VIVA data with uncertainties}
	
\begin{figure}
		\centering
		\includegraphics[width=1.0\columnwidth]{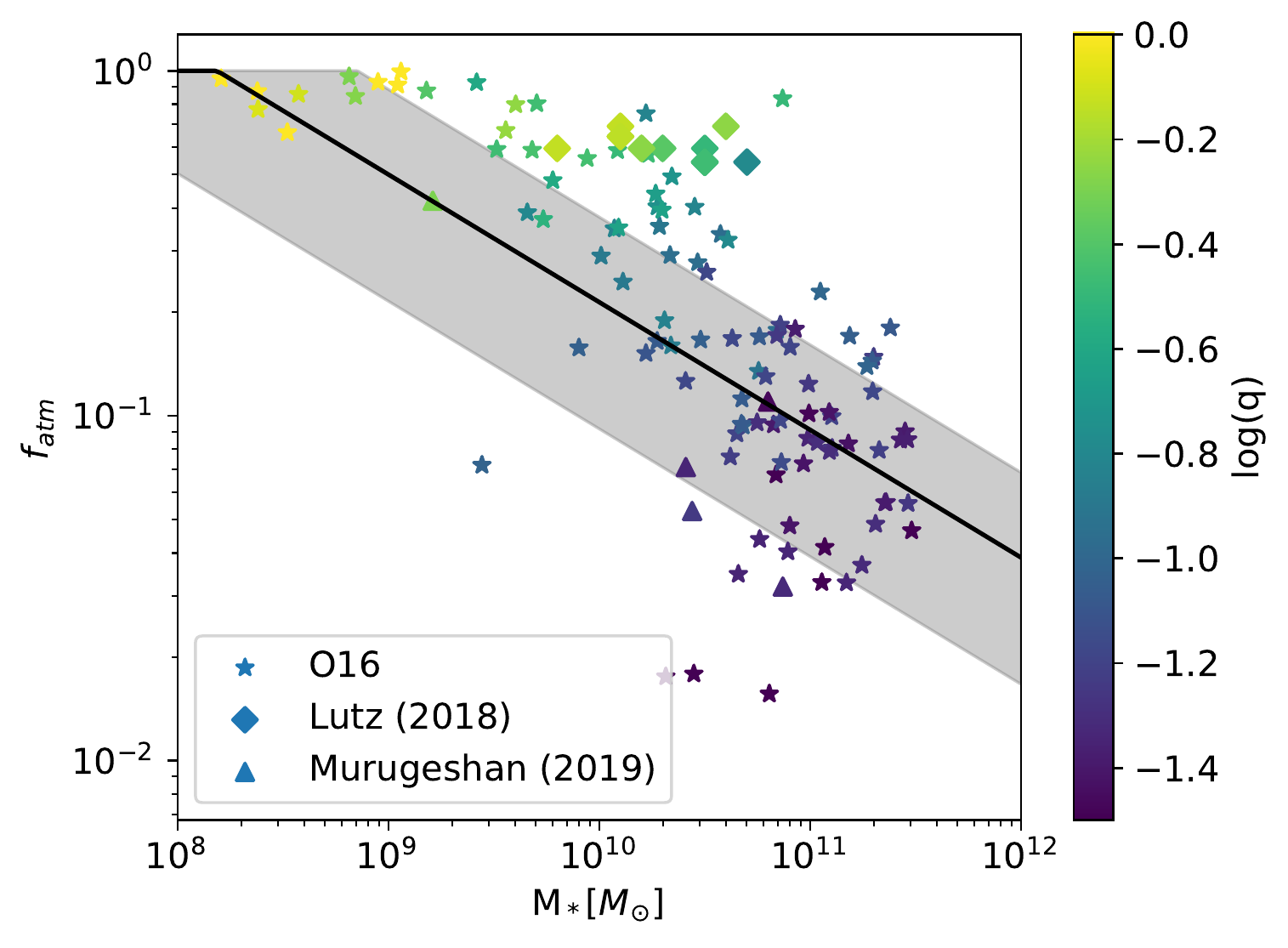}
		\caption{Disk mass vs. \HI gas fraction, colored by parameter $q$.  The samples are the same as in Figure~\ref{fig:definitionfq}. The black line shows the $M-\fatm$ relation derived from the analytic model in \citet{Obreschkow2016}. The shadow represents 68\% of the predicted scatter in the spin parameter.}
		\label{fig:Mfatm_j}
	\end{figure}

	Figure~\ref{fig:Mfatm_j} shows the $M-\fatm$ relation for disk field galaxies. The scatter of this relation is much larger than that of the $q-\fatm$ relation. The samples of \citet{Lutz2018} and \citet{Murugeshan2019} are \HI extreme for their mass, but for their stability parameter $q$, their \HI fractions are normal.
	
\begin{figure}
	\centering
	\includegraphics[width=1.0\columnwidth]{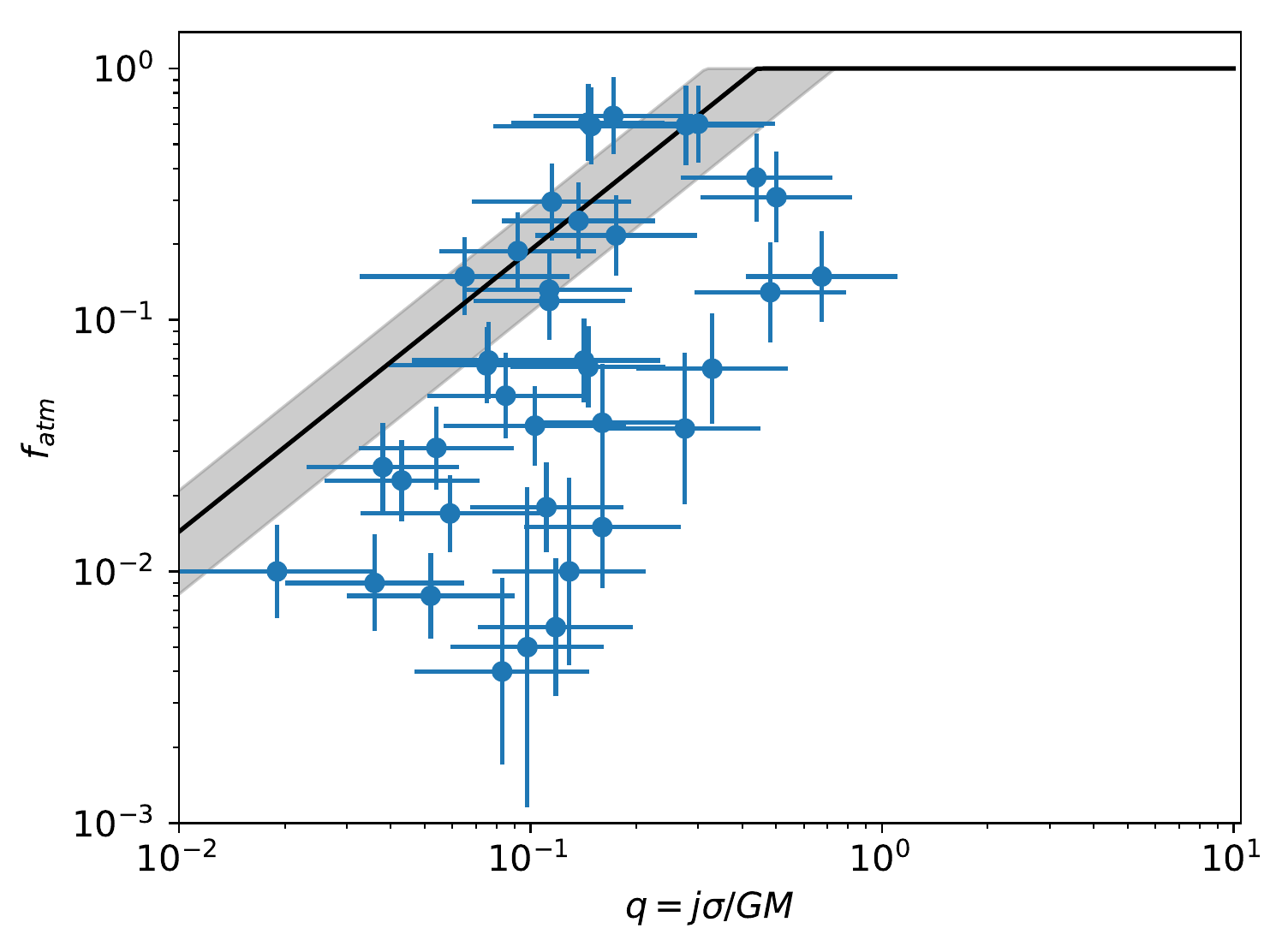}
	\caption{VIVA data in the $(q,\fatm)$-plane with log-normal uncertainties.}
	\label{fig:qf_error}
\end{figure}

	Figure~\ref{fig:qf_error} shows the $q$--$\fatm$ relation of the VIVA data with log-normal statistical uncertainties plotted as error bars. The uncertainties of $q$ combine the uncertainties of estimating $M_*$ \citep{Cortese2012}, $j$ derived from Section~\ref{VIVA} and a global 40\% uncertainty of the \HI velocity dispersion $\sigma$ \citep{Obreschkow2016}. The absolute uncertainties of $M_{\rm HI}$ and $M_{\rm H_2}$ are negligible compared to those of $M_*$. The uncertainty of $\fatm$ includes those of $M_*$ \citep{Cortese2012} and $M_{\rm HI}$ \citep{Chung2009}.
	
\bsp	
\label{lastpage}
\end{document}